\documentclass[letterpaper,english,prl,notitlepage,twocolumn,superscriptaddress]{revtex4-1}
\usepackage{verbatim}
\usepackage{amsmath}
\usepackage{amssymb}
\usepackage{graphicx}
\usepackage{dsfont}
\usepackage{soul}

\usepackage{natbib}
\setcitestyle{numbers}
\usepackage[colorlinks,citecolor=blue,urlcolor=blue]{hyperref}

\makeatletter

\usepackage{color}
\usepackage[usenames,dvipsnames,svgnames,table]{xcolor}
\usepackage{graphicx}
\usepackage{comment}
\usepackage[normalem]{ulem}
\definecolor{violet}{rgb}{0.58, 0.0, 0.83}
\definecolor{brown}{rgb}{0.7, 0.4, 0.4}

\newcommand{\add}[1]{#1}
\newcommand{\addcomment}[1]{}
\newcommand{\del}[1]{}
\newcommand{\delmath}[1]{}

\newcommand{\gae}{\lower 2pt \hbox{$\,
\buildrel{\scriptstyle >}\over {\scriptstyle \sim}\,$}}
\newcommand{\lae}{\lower 2pt \hbox{$\,
\buildrel{\scriptstyle <}\over {\scriptstyle \sim}\,$}}

\newcommand{\fref}[1]{Fig.~\ref{#1}}

\makeatother

\begin{document}
\title{Quantized Floquet topology with temporal noise}
\author{Christopher I. Timms}
\affiliation{Department of Physics, University of Texas at Dallas, Richardson, TX, USA}
\author{Lukas M. Sieberer}
\affiliation{Institute for Theoretical Physics, University of Innsbruck, 6020 Innsbruck, Austria}
\author{Michael H. Kolodrubetz}
\affiliation{Department of Physics, University of Texas at Dallas, Richardson, TX, USA}

\begin{abstract}
Time-periodic (Floquet) drive is a powerful method to engineer quantum phases of matter, including fundamentally non-equilibrium states that are impossible in static Hamiltonian systems. One characteristic example is the anomalous Floquet insulator, which exhibits topologically quantized chiral edge states similar to a Chern insulator, yet is amenable to bulk localization. We study the response of this topological system to time-dependent noise, which breaks the topologically protecting Floquet symmetry. Surprisingly, we find that the quantized response, given by partially filling the fermionic system and measuring charge pumped per cycle, remains quantized up to finite noise amplitude. We trace this robust topology to an interplay between diffusion and Pauli blocking of edge state decay, which we expect should be robust against interactions. \add{We determine the boundaries of the topological phase for a system with spatial disorder numerically through level statistics, and corroborate our results in the limit of vanishing disorder through an analytical Floquet superoperator approach. This approach suggests an interpretation of the state of the system as a non-Hermitian Floquet topological phase.} We comment on quantization of other topological responses in the absence of Floquet symmetry and potential experimental realizations.
\end{abstract}
\maketitle

\emph{Introduction} -- Periodic Floquet drive is an indispensable tool in engineered quantum systems \cite{Aidelsburger_2013, Wang_2013, Miyake_2013, Roushan2017, Boyers_2019,Oka_2019}. Recently, Floquet drive has enabled the realization of fundamentally non-equilibrium phases of matter, such as Floquet time crystals \cite{Keyserlingk2016, Else2016,Choi_2017,Zhang_2017,rovny2018observation,autti2018observation,Else2020,khemani2019brief} and Floquet symmetry-protected topological states (SPTs) \cite{chandran2014many,Nathan_2015,Roy2016,Keyserlingk2016a,Else2016a,potter2016classification,Roy2017,po2016chiral,Potter2017,Harper2017,Roy2017a,Po2017,Potter2018,Reiss2018}. A quintessential example of Floquet SPT is the anomalous Floquet-Anderson insulator (AFAI), which has topologically protected chiral edge states similar to a Chern insulator but with a fully localizable bulk  \cite{rudner2013anomalous,titumphysical,leykam2016anomalous,maczewsky2017observation}. Topologically protected transport in the AFAI can be measured via current flowing through the system \cite{titumphysical,kundu2020quantized}, magnetization density in a fully-filled patch within the bulk \cite{nathan2019anomalous}, or quantized transport of quantum information at the edge \cite{po2016chiral,PhysRevB.98.054309,fidkowski2019interacting}.

\begin{figure}[b]
	\includegraphics[width=\columnwidth]{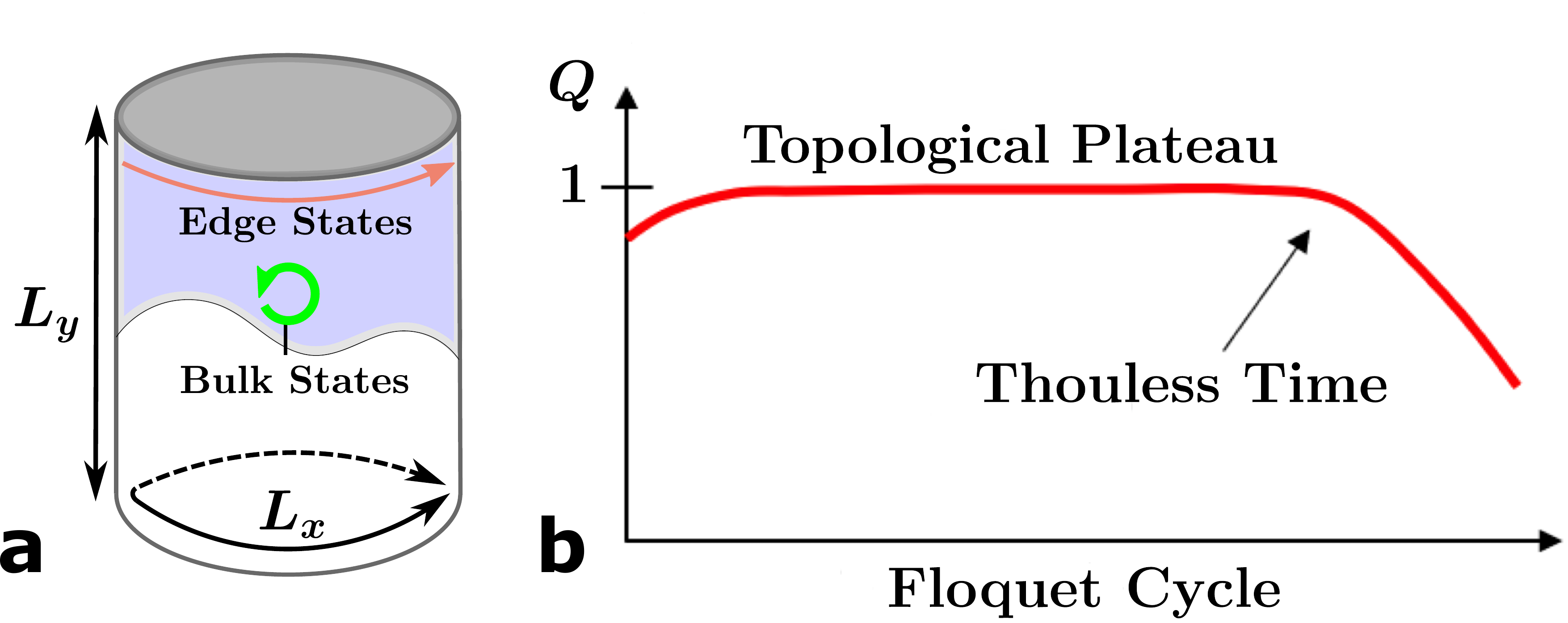}
	\caption{Illustration of quantized non-adiabatic pumping in the presence of noise. (a) The two-dimensional system is placed on a cylinder with the top half filled and bottom half empty, \add{and driven via a 5-step Floquet drive (\fref{Figure.Model})}. \add{Pumped charge} $Q$ around the cylinder per Floquet cycle is quantized without noise \add{due to topological edge states. The bulk states are localized, undergoing cyclotron-like orbits during each Floquet cycle (green arrow)}. Noise is added by disordering the timings of the 5-step drive\del{ (\fref{Figure.Model})}. (b) For weak noise, $Q$ goes to a topological plateau \add{after a non-universal short-time transient,} before decaying when the edge states start to depopulate at times of order the Thouless time.}
	\label{Figure.Intro}
\end{figure}

All of these non-equilibrium states are protected by discrete time-translation symmetry of the Floquet Hamiltonian, $H(t) = H(t+T)$, where $T=2\pi/\Omega$ is the driving period. In this Letter, we ask what happens to the AFAI upon breaking time-translation symmetry via time-dependent random noise. A similar question has been studied in the case of a Floquet SPT protected by chiral symmetry 
\cite{sieberer2018statistical,rieder2018localization}, where the authors found that edge states decay at a slow but finite rate set by diffusion. In this work, we instead find that for the two most realistic experimental protocols, namely bulk magnetization or current measurements in partially-filled samples \add{as illustrated in Fig.~\ref{Figure.Intro}a}, the topological response remains \emph{fully protected} over a time scale that diverges in the thermodynamic limit. We trace this topological protection back to Pauli blocking, which prevents diffusive loss of the topological edge state pumping up to approximately the Thouless time \add{as shown in ~\fref{Figure.Intro}b}. We argue that the results should hold for many-body localization as well as Anderson localization, and comment on the potential for experimental realization.

\begin{figure}[t]
    \includegraphics[width=\columnwidth]{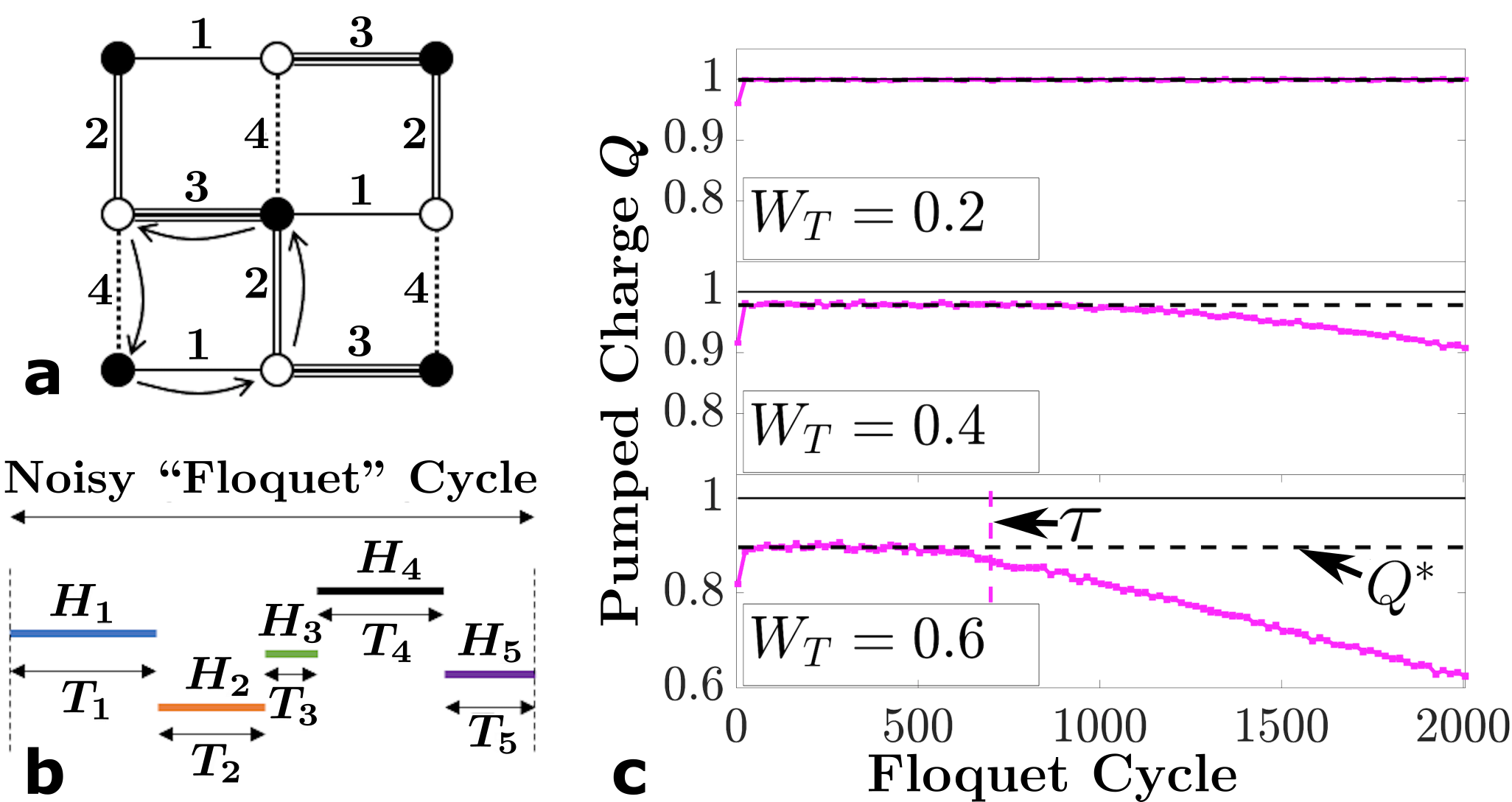}
    \caption{Noisy AFAI model. (a) First 4 steps of drive protocol. Hopping occurs on bonds labeled $1$ for $0<t<T_1$, on bonds labeled $2$ for $T_1 < t< T_1+T_2$, etc. \add{Filled (empty) circles are sites of sublattice $A$ ($B$).} (b) Noise is added by randomly changing the time over which the Hamiltonians are present, $T_\ell=T(1+\delta_\ell)/5$. The random noise $\delta_\ell\in[-W_T,W_T]$ is different for each ``Floquet'' cycle. (c) Charge pumped per Floquet cycle for 1D spatial disorder with $W=0.2$ and $L=100$\add{, averaged over spatial and temporal disorder}. Dashed lines show non-quantized plateau value. }
    \label{Figure.Model}
\end{figure}

\emph{Model} -- Throughout this paper, we study a single-particle model of the anomalous Floquet-Anderson insulator (AFAI) with time-dependent noise. We start from the original AFAI model \cite{titumphysical}, which involves a 5-step Floquet drive. The first four steps involve hopping between sites of the two sublattices. Specifically, for step $\ell\in \{1,2,3,4\}$, the Hamiltonian is $H_\ell=-J\sum_{\langle ij \rangle_\ell} c_i^\dagger c_j$, where \add{$c_j$ is the fermion annihilation operator on site $j$ and} $\langle ij \rangle_\ell$ indicates the bonds that are ``turned on'' during step $\ell$, as illustrated in \fref{Figure.Model}a. During step 5, a sublattice-dependent potential of strength $\Delta$ is applied: $H_5=\Delta \sum_{j} \eta_j c_j^\dagger c_j$, where $\eta_j=+1$ ($-1$) on the A (B) sublattice. Each Hamiltonian $H_\ell$ is present for time $T_\ell$, which in the absence of temporal order is just $T_\ell=T/5$. The hopping Hamiltonians $H_{1-4}$ are chosen such that, for the fine-tuned value 
$J=J_0\equiv 5 \Omega /4$, bulk electrons undergo a ``cyclotron'' orbit during each Floquet cycle and return to their original site, as illustrated in \add{\fref{Figure.Intro}a and } \fref{Figure.Model}a. A static chemical potential disorder is added throughout the cycle with Hamiltonian $H_\mathrm{dis}=\sum{\mu_j c_j^\dagger c_j}$, where each $\mu_j$ is uniformly sampled from the interval $[-W,W]$. Units are set by $\Omega=\hbar=1$, and we choose $\Delta = 0.4\Omega$ and $J=J_0=5\Omega/4$ throughout.

In this work, we modify the Hamiltonian by adding temporal disorder (noise). Explicitly, noise is introduced via random modification of the Floquet timing: $T_\ell = T(1+\delta_\ell)/5$, where $\delta_\ell \in [-W_T,W_T]$ is sampled \add{uniformly and} independently during each Floquet cycle \footnote{Note that one can equivalently keep each step fixed at $T/5$ and rescale the Hamiltonian by a factor $1+\delta_\ell$.}. Naively, one expects that noise will immediately destroy the Floquet topological phase, as it breaks the time periodicity \cite{sieberer2018statistical}. Yet, as we will show, the topological response remains robust against weak noise due to special properties of the AFAI's topological response.

In our numerics, we measure topologically protected non-adiabatic charge pumping for a cylinder of $L_x=2L$ and $L_y=L$ lattice sites \cite{titumphysical}. As shown in \fref{Figure.Intro}a, the system is initialized with one half of the cylindrical crystal filled with particles and the other half left empty. We measure the charge pumped \del{per} \add{during each} cycle,
\add{\begin{equation}
\label{ChargePumpedEquation}
Q = \int_{t_0}^{t_0+\tilde T} dt \langle\psi(t)|J_x|\psi(t)\rangle,
\end{equation}}
where $J_x$ is the current in the $x$-direction, $\tilde T = \sum_\ell T_\ell$ is the ``Floquet'' period appropriately modified by noise\add{, and $t_0$ is the time at the start of the cycle}. In the absence of temporal disorder, Titum et al.~\cite{titumphysical} demonstrated quantization of $Q$ in the presence of spatial disorder. One may think of this quantization as coming from the single filled edge state, which pumps $Q=1$ per cycle in the topological phase, while the localized bulk states do not carry current. In the presence of temporal disorder, the bulk states no longer remain localized; we now demonstrate how this affects $Q$.

\emph{One-dimensional disorder} -- Large two-dimensional (2D) lattices without translation symmetry are computationally challenging to simulate. Therefore, as a warmup problem in which we can address large system sizes, we begin by implementing one-dimensional (1D) spatial disorder in the $y$-direction, meaning that for site $j=(x,y)$, $\mu_j$ only depends on the $y$ position.

Some characteristic traces of $Q$ vs.~$t$ are shown in \fref{Figure.Model}c. For weak temporal and spatial disorder, the charge approaches a plateau value and remains nearly perfectly quantized up to more than 2000 drive cycles. As $W_T$ is increased, the plateau value of the pumped charge is no longer quantized and the pumped charge begins to decay at late times. To quantify this behavior, we define two quantities: the plateau value of pumped charge, $Q^{*}$, and the decay time scale, $\tau$.

\begin{figure}[b]
    \includegraphics[width=\columnwidth]{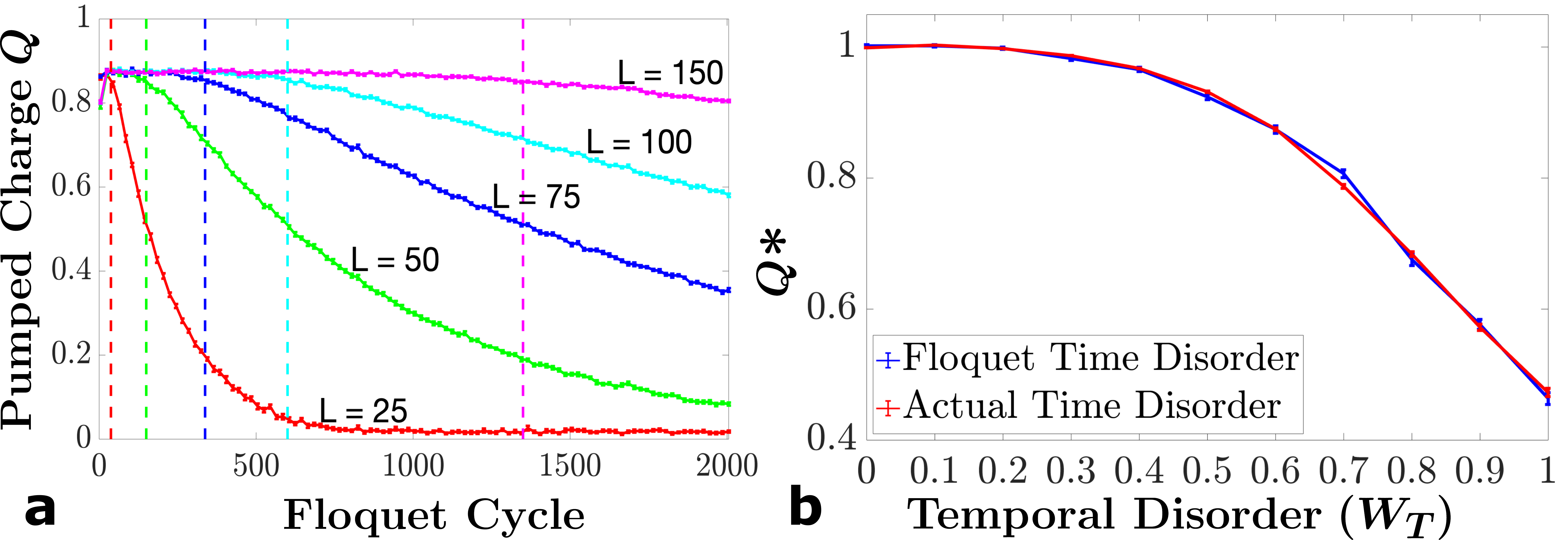}
  \caption{Finite size effects for 1D disorder. (a) System size dependence of $Q$ for $W=0.5$ and $W_T=0.6$. The dashed lines show times $\tau \sim L^2$, illustrating that the pumped charge begins to decay on \add{a time scale} of order the Thouless time, which is set by diffusion. (b) Comparison of plateau value $Q^\ast$ for actual time disorder and ``Floquet time disorder,'' in which the same random pattern of $\delta_\ell$ is repeated indefinitely. 
  Finite size effects have been removed by extrapolating to $L\to\infty$ using a linear fit to $Q^\ast$ versus $1/L$ at large $L$. \add{All data shown is averaged over spatial and temporal disorder.}}
    \label{Figure.finite_size_and_Qstar_1d_disorder}
\end{figure}

\begin{figure*}[t]
    \includegraphics[width=0.8\textwidth]{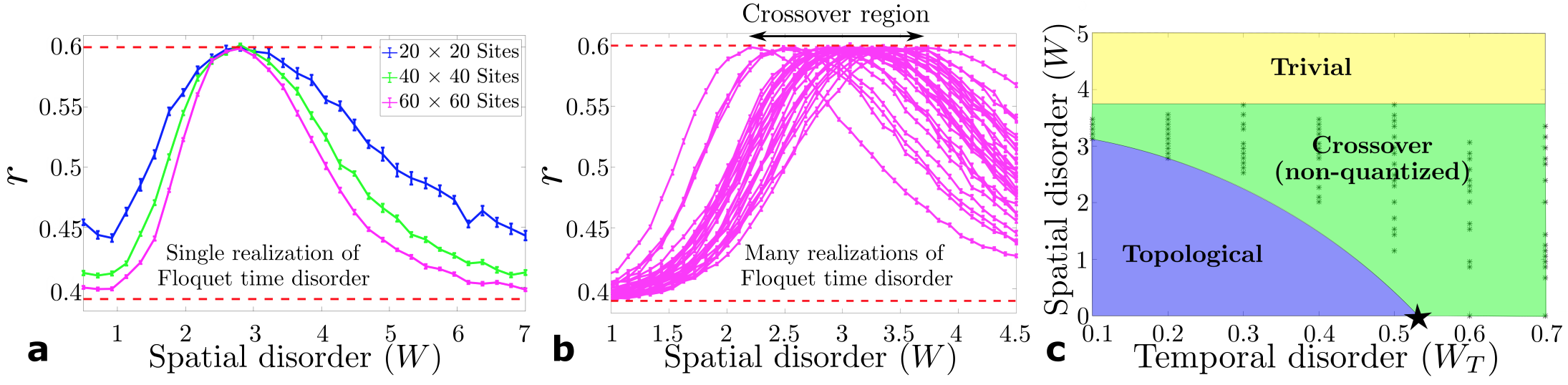}
  \caption{Topological phase diagram for 2D disorder in the presence of temporal noise. (a) Level spacing ratio $r$ averaged over spatial disorder for a single realization of Floquet temporal disorder with $W_T=0.3$. A clear peak is seen at $W_c\approx 2.8$, becoming increasingly sharp with increasing system size. We identify this as the critical point. (b) Data for system size of 60 $\times$ 60 sites with 30 different realizations of Floquet temporal disorder, showing that different realizations lead to different values of $W_c$. (c) Phase diagram obtained from peaks of $r$, plotted as black dots. \add{The value of $W_{T,c}$ for $W=0$, which is indicated with an asterisk, is obtained from gap closing of the noise-averaged Floquet superoperator (see SOM \cite{Supplement_PRL_Noisy_AFAI}).}}
    \label{Figure.2d_disorder}
\end{figure*}

The key to understanding these quantities is their dependence on system size $L$, shown in \fref{Figure.finite_size_and_Qstar_1d_disorder}\add{a}. We see that the plateau value $Q^{*}$ does not depend on system size, while $\tau$ increases sharply with system size. We have confirmed that this finite-size dependence of $\tau$ reflects the \add{known} fact that temporal disorder causes the particles of the system to undergo a diffusive random walk~\cite{rieder2018localization}. 
The consequence of this diffusion is that the sharp density edge separating the top and bottom half of the system spreads diffusively into a smooth \del{density dependence} \add{position dependence of the density}, until eventually the edge state starts to depopulate on a time scale of order the Thouless time, $t_\mathrm{D} =L^2/D$ with diffusion constant $D$. Since non-adiabatic charge pumping comes entirely from the edge state, the loss of edge state occupation corresponds to a loss of the signal in $Q$, and thus $\tau$ will be proportional to the Thouless time.

We can now draw two important conclusions about the system with one-dimensional disorder. First, the pumped charge reaches a plateau that eventually decays on a time scale $\tau \sim L^2$. Importantly, this implies that the plateau will be infinitely long-lived in the thermodynamic limit, where the topological phase is defined. 
Second, we learned that the plateau value $Q^\ast$ loses quantization as either spatial or temporal disorder are added\add{. We note in that this loss of quantization with $W$ is similar to the Floquet-Thouless energy pump~\cite{kolodrubetz2018topological},
where the conserved momentum $k_x$ in our system is replaced by an adiabatic pump parameter $\lambda$. Temporal disorder had not been studied earlier, but is causes a similar smooth reduct of $Q^\ast$ from its quantized value.}
We \del{postulate} \add{conjecture} that this physics is, in fact, exactly captured by that of the Floquet-Thouless energy pump. Specifically, we consider the behavior of a related Floquet system created by randomly sampling the times $T_1$, $T_2$, $\ldots$, $T_5$ as before, but then repeating this random sequence for each Floquet cycle. Such a Floquet system will achieve a plateau value $Q^\ast$ and then stay there \cite{kolodrubetz2018topological}, as there is no diffusion to prevent localization. We refer to this construction as ``Floquet time disorder.''

We compare the results of actual time disorder and Floquet time disorder in \fref{Figure.finite_size_and_Qstar_1d_disorder}b, showing that they match within error bars after extrapolation to the thermodynamic limit. Importantly, each realization of the Floquet time disorder can be analyzed in the language of the Floquet-Thouless energy pump, meaning that our non-topological response with time disorder is obtained by averaging over the non-quantized responses from the Floquet-Thouless energy pump. This explains why the response is not quantized, and provides a valuable method for defining (average) topology in this temporally disordered system.

\emph{Two-dimensional disorder} -- Having understood one-dimensional disorder, we can now make predictions for \add{the actual case of interest, namely} full two-dimensional disorder, in which $\mu_j$ is chosen independently for each site. The dependence of $\tau$ on system size \del{should} \add{will} be the same with 2D disorder, since \add{the mapping to a diffusive random walk} \del{diffusion} still applies. This means that the plateau value $Q^\ast$ should again be infinitely long-lived in the thermodynamic limit. However, a more interesting fact comes out of thinking about this plateau value. Unlike the case of 1D disorder, 2D disorder has a non-trivial topological phase (the AFAI) that survives to finite disorder, with a sharp transition from $Q=1$ to $Q=0$ at finite $W_c$ \cite{titumphysical}. Therefore, our analysis of 1D disorder implies that the non-trivial topological phase will also survive for weak Floquet time disorder, since this is a perturbative deformation of the original AFAI model. Given \add{that} \del{the equivalence of pumping for} time disorder and Floquet time disorder \add{demonstrate identical plateau values for $Q^\ast$ upon averaging over disorder configurations}, we thus predict that the AFAI is \emph{stable} to weak temporal noise. 

This intuition is confirmed numerically in \fref{Figure.2d_disorder} using the \add{a well-established technique introduced by} \del{same technique as} Titum et al. \cite{titumphysical}. Specifically, for a given realization of Floquet time disorder, we calculate the Floquet quasienergies $\epsilon^F_n$ and determine the statistics of their nearest-neighbor level spacings: $\Delta_n \equiv \epsilon^F_{n+1}-\epsilon^F_{n}$. We calculate the $r$-statistic \cite{PhysRevB.75.155111}:
\begin{equation}
    r_n = \mathrm{min}\left[\Delta_n,\Delta_{n+1}\right]/\mathrm{max}\left[\Delta_n,\Delta_{n+1}\right],
\end{equation}
whose \del{average value} \add{average over disorder and eigenstates $\langle r \rangle$} is a useful indicator of level repulsion. $\langle r \rangle$ converges to the Poisson value, $r_P \approx 0.39$, for localized systems that do not display level repulsion, and to the circular unitary ensemble (CUE) value, $r_C \approx 0.6$, for delocalized systems. In the present case of non-interacting particles, both the topologically non-trivial phase at low $W$ and the topologically trivial phase at high $W$ are localized, giving $r_P$. Right at the phase transition, the system delocalizes, creating a sharp peak with CUE level statistics. This peak was shown to be a sensitive indicator of the phase transition for the Floquet model \cite{titumphysical}, and we see this holds with Floquet time disorder as well (Figure \ref{Figure.2d_disorder}a) \footnote{Note that, for both $W_T=0$ and finite $W_T$, the plateau value $Q^\ast$ is not a sensitive indicator of the phase transition, requiring inaccesibly large system size in order to see a sharp transition}. Therefore, for a given realization of Floquet time disorder, we can obtain the critical disorder value $W_c$ by finding this peak.

There is one notable effect of Floquet time disorder, namely that different realizations of time disorder yield different values for this critical $W_c$, as seen in Figure \ref{Figure.2d_disorder}b. In other words, Floquet time disorder does not self average. This means that there is not a sharp transition from topologically non-trivial to trivial, but rather a topologically non-trivial phase for $W < W_{c,min}$, a topologically trivial phase for $W > W_{c,max}$, and a crossover region in between where the response is not quantized. The full phase diagram showing these three regions is plotted in \fref{Figure.2d_disorder}c, with best estimates for the phase transition lines $W_{c,min/max}$. \del{The topological phase survives up to a relatively large finite value $W_T \approx 0.6$.}

\add{\emph{Floquet superoperator approach} -- The topological transition can be obtained directly from the noise-averaged Floquet superoperator; an approach, which unlike that of the level spacing ratio, does not involve the auxiliary system with Floquet time disorder. The evolution of the density matrix $\rho$ of a single particle during a noisy Floquet cycle is described by a superoperator $\mathcal{U} \rho = U \rho U^{\dagger}$, where $U = e^{- i T_5 H_5} \dotsb e^{- i T_1 H_1}$. While spatial disorder can be incorporated into this superoperator, we consider a system with no disorder. Averaging over temporal noise, this becomes a non-unitary Floquet superoperator $\mathcal{F}$, whose eigenvalues lie within the unit circle on the complex plane \cite{sieberer2018statistical}. We analyze this superoperator in detail in the SOM \cite{Supplement_PRL_Noisy_AFAI}. We find that a gap closes on the real axis at $W_{T,c}=0.535$, which is indicated in \fref{Figure.2d_disorder}c with an astersik, suggesting a topological transition. Furthermore, one can define a generalized winding number for this superoperator, which ceases to be quantized for $W_T > W_{T,c}$ due to issues taking a branch cut along the real axis. This is consistent with the non-quantized crossover regime found earlier, and provides a readily generalizable, complementary perspective on our topological phase diagram. The Floquet superoperator analysis also allows us to define our topological system in the language of non-Hermitian Floquet topological SPTs~\cite{Lee2019_1}, which should be readily extensible to other systems and symmetry classes.}

\emph{Discussion} -- We have shown that the two-dimensional anomalous Floquet-Anderson insulator is stable to weak temporal noise. The argument involves constructing a related Floquet system for a given noise realization and then arguing that if each such realization is topological, then their noise-average\add{, which is given by the superoperator approach,} is topological as well. This argument should hold for other types of environmental noise, and therefore we \del{postulate} \add{expect} that the AFAI is stable to a wide class of weak dissipative couplings. Correlated noise would kill this argument, hence we leave generic non-Markovian baths for future work. \add{These responses may also be stable to quasiperiodic driving, which leads to a variety of interesting steady states in other contexts~\cite{Friedman2020a,Else2020b,Nandy2017}.}

While we numerically studied the topological response via charge pumping in a half-filled system, our arguments indicate that a similar story should hold for other proposed experimental measurements of the anomalous Floquet insulator \cite{nathan2017quantized,nathan2019anomalous,PhysRevB.98.054309}. For instance, topologically quantized magnetization for a filled region of linear size \del{$\ell$} \add{$l$}~\cite{nathan2017quantized} should hold up to time \del{$\tau \sim \ell^2$} \add{$\tau \sim l^2$} and remain measurable by the same protocols. This fact will be important in practical experimental realizations, as there are always finite noise sources -- such as laser fluctuations or spontaneous emission into lattice lasers -- that break the Floquet symmetry of the problem.

It has recently been argued that the AFAI is stable to interactions \cite{nathan2019anomalous}, and we suspect the same will be true in the presence of noise. An interesting question is how noise affects other topological invariants that have been identified in the AFAI \cite{nathan2019hierarchy}, which are also theoretically measurable. Finally, we speculate that similar ideas may be used to demonstrate stability in other Floquet topological phases, such as the Floquet topological superconductor, with possible implications for robust quantum information \add{processing} and computation~\cite{PhysRevLett.106.220402,PhysRevB.100.041102}.

\emph{Acknowledgments} -- We would like to acknowledge useful discussions with P. Titum and F. Nathan. This work was performed with support from the National Science Foundation through award number DMR-1945529 and the Welch Foundation through award number AT-2036-20200401. We used the computational resources of the Lonestar 5 cluster operated by the Texas Advanced Computing Center at the University of Texas at Austin and the Ganymede and Topo clusters operated by the University of Texas at Dallas' Cyberinfrastructure \& Research Services Department.

\bibliography{dissipative_afai}

\begin{thebibliography}{51}%
\makeatletter
\providecommand \@ifxundefined [1]{%
 \@ifx{#1\undefined}
}%
\providecommand \@ifnum [1]{%
 \ifnum #1\expandafter \@firstoftwo
 \else \expandafter \@secondoftwo
 \fi
}%
\providecommand \@ifx [1]{%
 \ifx #1\expandafter \@firstoftwo
 \else \expandafter \@secondoftwo
 \fi
}%
\providecommand \natexlab [1]{#1}%
\providecommand \enquote  [1]{``#1''}%
\providecommand \bibnamefont  [1]{#1}%
\providecommand \bibfnamefont [1]{#1}%
\providecommand \citenamefont [1]{#1}%
\providecommand \href@noop [0]{\@secondoftwo}%
\providecommand \href [0]{\begingroup \@sanitize@url \@href}%
\providecommand \@href[1]{\@@startlink{#1}\@@href}%
\providecommand \@@href[1]{\endgroup#1\@@endlink}%
\providecommand \@sanitize@url [0]{\catcode `\\12\catcode `\$12\catcode
  `\&12\catcode `\#12\catcode `\^12\catcode `\_12\catcode `\%12\relax}%
\providecommand \@@startlink[1]{}%
\providecommand \@@endlink[0]{}%
\providecommand \url  [0]{\begingroup\@sanitize@url \@url }%
\providecommand \@url [1]{\endgroup\@href {#1}{\urlprefix }}%
\providecommand \urlprefix  [0]{URL }%
\providecommand \Eprint [0]{\href }%
\providecommand \doibase [0]{http://dx.doi.org/}%
\providecommand \selectlanguage [0]{\@gobble}%
\providecommand \bibinfo  [0]{\@secondoftwo}%
\providecommand \bibfield  [0]{\@secondoftwo}%
\providecommand \translation [1]{[#1]}%
\providecommand \BibitemOpen [0]{}%
\providecommand \bibitemStop [0]{}%
\providecommand \bibitemNoStop [0]{.\EOS\space}%
\providecommand \EOS [0]{\spacefactor3000\relax}%
\providecommand \BibitemShut  [1]{\csname bibitem#1\endcsname}%
\let\auto@bib@innerbib\@empty
\bibitem [{\citenamefont {Aidelsburger}\ \emph {et~al.}(2013)\citenamefont
  {Aidelsburger}, \citenamefont {Atala}, \citenamefont {Lohse}, \citenamefont
  {Barreiro}, \citenamefont {Paredes},\ and\ \citenamefont
  {Bloch}}]{Aidelsburger_2013}%
  \BibitemOpen
  \bibfield  {author} {\bibinfo {author} {\bibfnamefont {M.}~\bibnamefont
  {Aidelsburger}}, \bibinfo {author} {\bibfnamefont {M.}~\bibnamefont {Atala}},
  \bibinfo {author} {\bibfnamefont {M.}~\bibnamefont {Lohse}}, \bibinfo
  {author} {\bibfnamefont {J.~T.}\ \bibnamefont {Barreiro}}, \bibinfo {author}
  {\bibfnamefont {B.}~\bibnamefont {Paredes}}, \ and\ \bibinfo {author}
  {\bibfnamefont {I.}~\bibnamefont {Bloch}},\ }\href@noop {} {\bibfield
  {journal} {\bibinfo  {journal} {Phys. Rev. Lett.}\ }\textbf {\bibinfo
  {volume} {111}},\ \bibinfo {pages} {185301} (\bibinfo {year}
  {2013})}\BibitemShut {NoStop}%
\bibitem [{\citenamefont {Wang}\ \emph {et~al.}(2013)\citenamefont {Wang},
  \citenamefont {Steinberg}, \citenamefont {Jarillo-Herrero},\ and\
  \citenamefont {Gedik}}]{Wang_2013}%
  \BibitemOpen
  \bibfield  {author} {\bibinfo {author} {\bibfnamefont {Y.~H.}\ \bibnamefont
  {Wang}}, \bibinfo {author} {\bibfnamefont {H.}~\bibnamefont {Steinberg}},
  \bibinfo {author} {\bibfnamefont {P.}~\bibnamefont {Jarillo-Herrero}}, \ and\
  \bibinfo {author} {\bibfnamefont {N.}~\bibnamefont {Gedik}},\ }\href@noop {}
  {\bibfield  {journal} {\bibinfo  {journal} {Science}\ }\textbf {\bibinfo
  {volume} {342}},\ \bibinfo {pages} {453} (\bibinfo {year}
  {2013})}\BibitemShut {NoStop}%
\bibitem [{\citenamefont {Miyake}\ \emph {et~al.}(2013)\citenamefont {Miyake},
  \citenamefont {Siviloglou}, \citenamefont {Kennedy}, \citenamefont {Burton},\
  and\ \citenamefont {Ketterle}}]{Miyake_2013}%
  \BibitemOpen
  \bibfield  {author} {\bibinfo {author} {\bibfnamefont {H.}~\bibnamefont
  {Miyake}}, \bibinfo {author} {\bibfnamefont {G.~A.}\ \bibnamefont
  {Siviloglou}}, \bibinfo {author} {\bibfnamefont {C.~J.}\ \bibnamefont
  {Kennedy}}, \bibinfo {author} {\bibfnamefont {W.~C.}\ \bibnamefont {Burton}},
  \ and\ \bibinfo {author} {\bibfnamefont {W.}~\bibnamefont {Ketterle}},\
  }\href@noop {} {\bibfield  {journal} {\bibinfo  {journal} {Phys. Rev. Lett.}\
  }\textbf {\bibinfo {volume} {111}},\ \bibinfo {pages} {185302} (\bibinfo
  {year} {2013})}\BibitemShut {NoStop}%
\bibitem [{\citenamefont {Roushan}\ \emph {et~al.}(2017)\citenamefont
  {Roushan}, \citenamefont {Neill}, \citenamefont {Megrant}, \citenamefont
  {Chen}, \citenamefont {Babbush}, \citenamefont {Barends}, \citenamefont
  {Campbell}, \citenamefont {Chen}, \citenamefont {Chiaro},\ and\ \citenamefont
  {Dunsworth}}]{Roushan2017}%
  \BibitemOpen
  \bibfield  {author} {\bibinfo {author} {\bibfnamefont {P.}~\bibnamefont
  {Roushan}}, \bibinfo {author} {\bibfnamefont {C.}~\bibnamefont {Neill}},
  \bibinfo {author} {\bibfnamefont {A.}~\bibnamefont {Megrant}}, \bibinfo
  {author} {\bibfnamefont {Y.}~\bibnamefont {Chen}}, \bibinfo {author}
  {\bibfnamefont {R.}~\bibnamefont {Babbush}}, \bibinfo {author} {\bibfnamefont
  {R.}~\bibnamefont {Barends}}, \bibinfo {author} {\bibfnamefont
  {B.}~\bibnamefont {Campbell}}, \bibinfo {author} {\bibfnamefont
  {Z.}~\bibnamefont {Chen}}, \bibinfo {author} {\bibfnamefont {B.}~\bibnamefont
  {Chiaro}}, \ and\ \bibinfo {author} {\bibfnamefont {A.}~\bibnamefont
  {Dunsworth}},\ }\href@noop {} {\bibfield  {journal} {\bibinfo  {journal}
  {Nature Physics}\ }\textbf {\bibinfo {volume} {13}},\ \bibinfo {pages} {146}
  (\bibinfo {year} {2017})}\BibitemShut {NoStop}%
\bibitem [{\citenamefont {Boyers}\ \emph {et~al.}(2019)\citenamefont {Boyers},
  \citenamefont {Pandey}, \citenamefont {Campbell}, \citenamefont
  {Polkovnikov}, \citenamefont {Sels},\ and\ \citenamefont
  {Sushkov}}]{Boyers_2019}%
  \BibitemOpen
  \bibfield  {author} {\bibinfo {author} {\bibfnamefont {E.}~\bibnamefont
  {Boyers}}, \bibinfo {author} {\bibfnamefont {M.}~\bibnamefont {Pandey}},
  \bibinfo {author} {\bibfnamefont {D.~K.}\ \bibnamefont {Campbell}}, \bibinfo
  {author} {\bibfnamefont {A.}~\bibnamefont {Polkovnikov}}, \bibinfo {author}
  {\bibfnamefont {D.}~\bibnamefont {Sels}}, \ and\ \bibinfo {author}
  {\bibfnamefont {A.~O.}\ \bibnamefont {Sushkov}},\ }\href@noop {} {\bibfield
  {journal} {\bibinfo  {journal} {Phys. Rev. A}\ }\textbf {\bibinfo {volume}
  {100}},\ \bibinfo {pages} {012341} (\bibinfo {year} {2019})}\BibitemShut
  {NoStop}%
\bibitem [{\citenamefont {Oka}\ and\ \citenamefont
  {Kitamura}(2019)}]{Oka_2019}%
  \BibitemOpen
  \bibfield  {author} {\bibinfo {author} {\bibfnamefont {T.}~\bibnamefont
  {Oka}}\ and\ \bibinfo {author} {\bibfnamefont {S.}~\bibnamefont {Kitamura}},\
  }\href@noop {} {\bibfield  {journal} {\bibinfo  {journal} {Annual Review of
  Condensed Matter Physics}\ }\textbf {\bibinfo {volume} {10}},\ \bibinfo
  {pages} {387} (\bibinfo {year} {2019})}\BibitemShut {NoStop}%
\bibitem [{\citenamefont {von Keyserlingk}\ and\ \citenamefont
  {Sondhi}(2016{\natexlab{a}})}]{Keyserlingk2016}%
  \BibitemOpen
  \bibfield  {author} {\bibinfo {author} {\bibfnamefont {C.~W.}\ \bibnamefont
  {von Keyserlingk}}\ and\ \bibinfo {author} {\bibfnamefont {S.~L.}\
  \bibnamefont {Sondhi}},\ }\href@noop {} {\bibfield  {journal} {\bibinfo
  {journal} {Phys. Rev. B}\ }\textbf {\bibinfo {volume} {93}},\ \bibinfo
  {pages} {245146} (\bibinfo {year} {2016}{\natexlab{a}})}\BibitemShut
  {NoStop}%
\bibitem [{\citenamefont {Else}\ \emph {et~al.}(2016)\citenamefont {Else},
  \citenamefont {Bauer},\ and\ \citenamefont {Nayak}}]{Else2016}%
  \BibitemOpen
  \bibfield  {author} {\bibinfo {author} {\bibfnamefont {D.~V.}\ \bibnamefont
  {Else}}, \bibinfo {author} {\bibfnamefont {B.}~\bibnamefont {Bauer}}, \ and\
  \bibinfo {author} {\bibfnamefont {C.}~\bibnamefont {Nayak}},\ }\href@noop {}
  {\bibfield  {journal} {\bibinfo  {journal} {Phys. Rev. Lett.}\ }\textbf
  {\bibinfo {volume} {117}},\ \bibinfo {pages} {090402} (\bibinfo {year}
  {2016})}\BibitemShut {NoStop}%
\bibitem [{\citenamefont {Choi}\ \emph {et~al.}(2017)\citenamefont {Choi},
  \citenamefont {Choi}, \citenamefont {Landig}, \citenamefont {Kucsko},
  \citenamefont {Zhou}, \citenamefont {Isoya}, \citenamefont {Jelezko},
  \citenamefont {Onoda}, \citenamefont {Sumiya},\ and\ \citenamefont
  {Khemani}}]{Choi_2017}%
  \BibitemOpen
  \bibfield  {author} {\bibinfo {author} {\bibfnamefont {S.}~\bibnamefont
  {Choi}}, \bibinfo {author} {\bibfnamefont {J.}~\bibnamefont {Choi}}, \bibinfo
  {author} {\bibfnamefont {R.}~\bibnamefont {Landig}}, \bibinfo {author}
  {\bibfnamefont {G.}~\bibnamefont {Kucsko}}, \bibinfo {author} {\bibfnamefont
  {H.}~\bibnamefont {Zhou}}, \bibinfo {author} {\bibfnamefont {J.}~\bibnamefont
  {Isoya}}, \bibinfo {author} {\bibfnamefont {F.}~\bibnamefont {Jelezko}},
  \bibinfo {author} {\bibfnamefont {S.}~\bibnamefont {Onoda}}, \bibinfo
  {author} {\bibfnamefont {H.}~\bibnamefont {Sumiya}}, \ and\ \bibinfo {author}
  {\bibfnamefont {V.}~\bibnamefont {Khemani}},\ }\href@noop {} {\bibfield
  {journal} {\bibinfo  {journal} {Nature}\ }\textbf {\bibinfo {volume} {543}},\
  \bibinfo {pages} {221} (\bibinfo {year} {2017})}\BibitemShut {NoStop}%
\bibitem [{\citenamefont {Zhang}\ \emph {et~al.}(2017)\citenamefont {Zhang},
  \citenamefont {Hess}, \citenamefont {Kyprianidis}, \citenamefont {Becker},
  \citenamefont {Lee}, \citenamefont {Smith}, \citenamefont {Pagano},
  \citenamefont {Potirniche}, \citenamefont {Potter},\ and\ \citenamefont
  {Vishwanath}}]{Zhang_2017}%
  \BibitemOpen
  \bibfield  {author} {\bibinfo {author} {\bibfnamefont {J.}~\bibnamefont
  {Zhang}}, \bibinfo {author} {\bibfnamefont {P.~W.}\ \bibnamefont {Hess}},
  \bibinfo {author} {\bibfnamefont {A.}~\bibnamefont {Kyprianidis}}, \bibinfo
  {author} {\bibfnamefont {P.}~\bibnamefont {Becker}}, \bibinfo {author}
  {\bibfnamefont {A.}~\bibnamefont {Lee}}, \bibinfo {author} {\bibfnamefont
  {J.}~\bibnamefont {Smith}}, \bibinfo {author} {\bibfnamefont
  {G.}~\bibnamefont {Pagano}}, \bibinfo {author} {\bibfnamefont {I.-D.}\
  \bibnamefont {Potirniche}}, \bibinfo {author} {\bibfnamefont {A.~C.}\
  \bibnamefont {Potter}}, \ and\ \bibinfo {author} {\bibfnamefont
  {A.}~\bibnamefont {Vishwanath}},\ }\href@noop {} {\bibfield  {journal}
  {\bibinfo  {journal} {Nature}\ }\textbf {\bibinfo {volume} {543}},\ \bibinfo
  {pages} {217} (\bibinfo {year} {2017})}\BibitemShut {NoStop}%
\bibitem [{\citenamefont {Rovny}\ \emph {et~al.}(2018)\citenamefont {Rovny},
  \citenamefont {Blum},\ and\ \citenamefont {Barrett}}]{rovny2018observation}%
  \BibitemOpen
  \bibfield  {author} {\bibinfo {author} {\bibfnamefont {J.}~\bibnamefont
  {Rovny}}, \bibinfo {author} {\bibfnamefont {R.~L.}\ \bibnamefont {Blum}}, \
  and\ \bibinfo {author} {\bibfnamefont {S.~E.}\ \bibnamefont {Barrett}},\
  }\href@noop {} {\bibfield  {journal} {\bibinfo  {journal} {Phys. Rev. Lett.}\
  }\textbf {\bibinfo {volume} {120}},\ \bibinfo {pages} {180603} (\bibinfo
  {year} {2018})}\BibitemShut {NoStop}%
\bibitem [{\citenamefont {Autti}\ \emph {et~al.}(2018)\citenamefont {Autti},
  \citenamefont {Eltsov},\ and\ \citenamefont
  {Volovik}}]{autti2018observation}%
  \BibitemOpen
  \bibfield  {author} {\bibinfo {author} {\bibfnamefont {S.}~\bibnamefont
  {Autti}}, \bibinfo {author} {\bibfnamefont {V.~B.}\ \bibnamefont {Eltsov}}, \
  and\ \bibinfo {author} {\bibfnamefont {G.~E.}\ \bibnamefont {Volovik}},\
  }\href@noop {} {\bibfield  {journal} {\bibinfo  {journal} {Phys. Rev. Lett.}\
  }\textbf {\bibinfo {volume} {120}},\ \bibinfo {pages} {215301} (\bibinfo
  {year} {2018})}\BibitemShut {NoStop}%
\bibitem [{\citenamefont {Else}\ \emph
  {et~al.}(2020{\natexlab{a}})\citenamefont {Else}, \citenamefont {Monroe},
  \citenamefont {Nayak},\ and\ \citenamefont {Yao}}]{Else2020}%
  \BibitemOpen
  \bibfield  {author} {\bibinfo {author} {\bibfnamefont {D.~V.}\ \bibnamefont
  {Else}}, \bibinfo {author} {\bibfnamefont {C.}~\bibnamefont {Monroe}},
  \bibinfo {author} {\bibfnamefont {C.}~\bibnamefont {Nayak}}, \ and\ \bibinfo
  {author} {\bibfnamefont {N.~Y.}\ \bibnamefont {Yao}},\ }\href@noop {}
  {\bibfield  {journal} {\bibinfo  {journal} {Annual Review of Condensed Matter
  Physics}\ }\textbf {\bibinfo {volume} {11}},\ \bibinfo {pages} {467}
  (\bibinfo {year} {2020}{\natexlab{a}})}\BibitemShut {NoStop}%
\bibitem [{\citenamefont {Khemani}\ \emph {et~al.}(2019)\citenamefont
  {Khemani}, \citenamefont {Moessner},\ and\ \citenamefont
  {Sondhi}}]{khemani2019brief}%
  \BibitemOpen
  \bibfield  {author} {\bibinfo {author} {\bibfnamefont {V.}~\bibnamefont
  {Khemani}}, \bibinfo {author} {\bibfnamefont {R.}~\bibnamefont {Moessner}}, \
  and\ \bibinfo {author} {\bibfnamefont {S.~L.}\ \bibnamefont {Sondhi}},\
  }\href@noop {} {\bibfield  {journal} {\bibinfo  {journal} {arXiv preprint
  arXiv:1910.10745}\ } (\bibinfo {year} {2019})}\BibitemShut {NoStop}%
\bibitem [{\citenamefont {Chandran}\ \emph {et~al.}(2014)\citenamefont
  {Chandran}, \citenamefont {Khemani}, \citenamefont {Laumann},\ and\
  \citenamefont {Sondhi}}]{chandran2014many}%
  \BibitemOpen
  \bibfield  {author} {\bibinfo {author} {\bibfnamefont {A.}~\bibnamefont
  {Chandran}}, \bibinfo {author} {\bibfnamefont {V.}~\bibnamefont {Khemani}},
  \bibinfo {author} {\bibfnamefont {C.~R.}\ \bibnamefont {Laumann}}, \ and\
  \bibinfo {author} {\bibfnamefont {S.~L.}\ \bibnamefont {Sondhi}},\
  }\href@noop {} {\bibfield  {journal} {\bibinfo  {journal} {Phys. Rev. B}\
  }\textbf {\bibinfo {volume} {89}},\ \bibinfo {pages} {144201} (\bibinfo
  {year} {2014})}\BibitemShut {NoStop}%
\bibitem [{\citenamefont {Nathan}\ and\ \citenamefont
  {Rudner}(2015)}]{Nathan_2015}%
  \BibitemOpen
  \bibfield  {author} {\bibinfo {author} {\bibfnamefont {F.}~\bibnamefont
  {Nathan}}\ and\ \bibinfo {author} {\bibfnamefont {M.~S.}\ \bibnamefont
  {Rudner}},\ }\href@noop {} {\bibfield  {journal} {\bibinfo  {journal} {New
  Journal of Physics}\ }\textbf {\bibinfo {volume} {17}},\ \bibinfo {pages}
  {125014} (\bibinfo {year} {2015})}\BibitemShut {NoStop}%
\bibitem [{\citenamefont {Roy}\ and\ \citenamefont {Harper}(2016)}]{Roy2016}%
  \BibitemOpen
  \bibfield  {author} {\bibinfo {author} {\bibfnamefont {R.}~\bibnamefont
  {Roy}}\ and\ \bibinfo {author} {\bibfnamefont {F.}~\bibnamefont {Harper}},\
  }\href@noop {} {\bibfield  {journal} {\bibinfo  {journal} {Phys. Rev. B}\
  }\textbf {\bibinfo {volume} {94}},\ \bibinfo {pages} {125105} (\bibinfo
  {year} {2016})}\BibitemShut {NoStop}%
\bibitem [{\citenamefont {von Keyserlingk}\ and\ \citenamefont
  {Sondhi}(2016{\natexlab{b}})}]{Keyserlingk2016a}%
  \BibitemOpen
  \bibfield  {author} {\bibinfo {author} {\bibfnamefont {C.~W.}\ \bibnamefont
  {von Keyserlingk}}\ and\ \bibinfo {author} {\bibfnamefont {S.~L.}\
  \bibnamefont {Sondhi}},\ }\href@noop {} {\bibfield  {journal} {\bibinfo
  {journal} {Phys. Rev. B}\ }\textbf {\bibinfo {volume} {93}},\ \bibinfo
  {pages} {245145} (\bibinfo {year} {2016}{\natexlab{b}})}\BibitemShut
  {NoStop}%
\bibitem [{\citenamefont {Else}\ and\ \citenamefont {Nayak}(2016)}]{Else2016a}%
  \BibitemOpen
  \bibfield  {author} {\bibinfo {author} {\bibfnamefont {D.~V.}\ \bibnamefont
  {Else}}\ and\ \bibinfo {author} {\bibfnamefont {C.}~\bibnamefont {Nayak}},\
  }\href@noop {} {\bibfield  {journal} {\bibinfo  {journal} {Phys. Rev. B}\
  }\textbf {\bibinfo {volume} {93}},\ \bibinfo {pages} {201103} (\bibinfo
  {year} {2016})}\BibitemShut {NoStop}%
\bibitem [{\citenamefont {Potter}\ \emph {et~al.}(2016)\citenamefont {Potter},
  \citenamefont {Morimoto},\ and\ \citenamefont
  {Vishwanath}}]{potter2016classification}%
  \BibitemOpen
  \bibfield  {author} {\bibinfo {author} {\bibfnamefont {A.~C.}\ \bibnamefont
  {Potter}}, \bibinfo {author} {\bibfnamefont {T.}~\bibnamefont {Morimoto}}, \
  and\ \bibinfo {author} {\bibfnamefont {A.}~\bibnamefont {Vishwanath}},\
  }\href@noop {} {\bibfield  {journal} {\bibinfo  {journal} {Phys. Rev. X}\
  }\textbf {\bibinfo {volume} {6}},\ \bibinfo {pages} {041001} (\bibinfo {year}
  {2016})}\BibitemShut {NoStop}%
\bibitem [{\citenamefont {Roy}\ and\ \citenamefont
  {Harper}(2017{\natexlab{a}})}]{Roy2017}%
  \BibitemOpen
  \bibfield  {author} {\bibinfo {author} {\bibfnamefont {R.}~\bibnamefont
  {Roy}}\ and\ \bibinfo {author} {\bibfnamefont {F.}~\bibnamefont {Harper}},\
  }\href@noop {} {\bibfield  {journal} {\bibinfo  {journal} {Phys. Rev. B}\
  }\textbf {\bibinfo {volume} {96}},\ \bibinfo {pages} {155118} (\bibinfo
  {year} {2017}{\natexlab{a}})}\BibitemShut {NoStop}%
\bibitem [{\citenamefont {Po}\ \emph {et~al.}(2016)\citenamefont {Po},
  \citenamefont {Fidkowski}, \citenamefont {Morimoto}, \citenamefont {Potter},\
  and\ \citenamefont {Vishwanath}}]{po2016chiral}%
  \BibitemOpen
  \bibfield  {author} {\bibinfo {author} {\bibfnamefont {H.~C.}\ \bibnamefont
  {Po}}, \bibinfo {author} {\bibfnamefont {L.}~\bibnamefont {Fidkowski}},
  \bibinfo {author} {\bibfnamefont {T.}~\bibnamefont {Morimoto}}, \bibinfo
  {author} {\bibfnamefont {A.~C.}\ \bibnamefont {Potter}}, \ and\ \bibinfo
  {author} {\bibfnamefont {A.}~\bibnamefont {Vishwanath}},\ }\href@noop {}
  {\bibfield  {journal} {\bibinfo  {journal} {Phys. Rev. X}\ }\textbf {\bibinfo
  {volume} {6}},\ \bibinfo {pages} {041070} (\bibinfo {year}
  {2016})}\BibitemShut {NoStop}%
\bibitem [{\citenamefont {Potter}\ and\ \citenamefont
  {Morimoto}(2017)}]{Potter2017}%
  \BibitemOpen
  \bibfield  {author} {\bibinfo {author} {\bibfnamefont {A.~C.}\ \bibnamefont
  {Potter}}\ and\ \bibinfo {author} {\bibfnamefont {T.}~\bibnamefont
  {Morimoto}},\ }\href@noop {} {\bibfield  {journal} {\bibinfo  {journal}
  {Phys. Rev. B}\ }\textbf {\bibinfo {volume} {95}},\ \bibinfo {pages} {155126}
  (\bibinfo {year} {2017})}\BibitemShut {NoStop}%
\bibitem [{\citenamefont {Harper}\ and\ \citenamefont
  {Roy}(2017)}]{Harper2017}%
  \BibitemOpen
  \bibfield  {author} {\bibinfo {author} {\bibfnamefont {F.}~\bibnamefont
  {Harper}}\ and\ \bibinfo {author} {\bibfnamefont {R.}~\bibnamefont {Roy}},\
  }\href@noop {} {\bibfield  {journal} {\bibinfo  {journal} {Phys. Rev. Lett.}\
  }\textbf {\bibinfo {volume} {118}},\ \bibinfo {pages} {115301} (\bibinfo
  {year} {2017})}\BibitemShut {NoStop}%
\bibitem [{\citenamefont {Roy}\ and\ \citenamefont
  {Harper}(2017{\natexlab{b}})}]{Roy2017a}%
  \BibitemOpen
  \bibfield  {author} {\bibinfo {author} {\bibfnamefont {R.}~\bibnamefont
  {Roy}}\ and\ \bibinfo {author} {\bibfnamefont {F.}~\bibnamefont {Harper}},\
  }\href@noop {} {\bibfield  {journal} {\bibinfo  {journal} {Phys. Rev. B}\
  }\textbf {\bibinfo {volume} {95}},\ \bibinfo {pages} {195128} (\bibinfo
  {year} {2017}{\natexlab{b}})}\BibitemShut {NoStop}%
\bibitem [{\citenamefont {Po}\ \emph {et~al.}(2017)\citenamefont {Po},
  \citenamefont {Fidkowski}, \citenamefont {Vishwanath},\ and\ \citenamefont
  {Potter}}]{Po2017}%
  \BibitemOpen
  \bibfield  {author} {\bibinfo {author} {\bibfnamefont {H.~C.}\ \bibnamefont
  {Po}}, \bibinfo {author} {\bibfnamefont {L.}~\bibnamefont {Fidkowski}},
  \bibinfo {author} {\bibfnamefont {A.}~\bibnamefont {Vishwanath}}, \ and\
  \bibinfo {author} {\bibfnamefont {A.~C.}\ \bibnamefont {Potter}},\
  }\href@noop {} {\bibfield  {journal} {\bibinfo  {journal} {Phys. Rev. B}\
  }\textbf {\bibinfo {volume} {96}},\ \bibinfo {pages} {245116} (\bibinfo
  {year} {2017})}\BibitemShut {NoStop}%
\bibitem [{\citenamefont {Potter}\ \emph {et~al.}(2018)\citenamefont {Potter},
  \citenamefont {Vishwanath},\ and\ \citenamefont {Fidkowski}}]{Potter2018}%
  \BibitemOpen
  \bibfield  {author} {\bibinfo {author} {\bibfnamefont {A.~C.}\ \bibnamefont
  {Potter}}, \bibinfo {author} {\bibfnamefont {A.}~\bibnamefont {Vishwanath}},
  \ and\ \bibinfo {author} {\bibfnamefont {L.}~\bibnamefont {Fidkowski}},\
  }\href@noop {} {\bibfield  {journal} {\bibinfo  {journal} {Phys. Rev. B}\
  }\textbf {\bibinfo {volume} {97}},\ \bibinfo {pages} {245106} (\bibinfo
  {year} {2018})}\BibitemShut {NoStop}%
\bibitem [{\citenamefont {Reiss}\ \emph {et~al.}(2018)\citenamefont {Reiss},
  \citenamefont {Harper},\ and\ \citenamefont {Roy}}]{Reiss2018}%
  \BibitemOpen
  \bibfield  {author} {\bibinfo {author} {\bibfnamefont {D.}~\bibnamefont
  {Reiss}}, \bibinfo {author} {\bibfnamefont {F.}~\bibnamefont {Harper}}, \
  and\ \bibinfo {author} {\bibfnamefont {R.}~\bibnamefont {Roy}},\ }\href@noop
  {} {\bibfield  {journal} {\bibinfo  {journal} {Phys. Rev. B}\ }\textbf
  {\bibinfo {volume} {98}},\ \bibinfo {pages} {045127} (\bibinfo {year}
  {2018})}\BibitemShut {NoStop}%
\bibitem [{\citenamefont {Rudner}\ \emph {et~al.}(2013)\citenamefont {Rudner},
  \citenamefont {Lindner}, \citenamefont {Berg},\ and\ \citenamefont
  {Levin}}]{rudner2013anomalous}%
  \BibitemOpen
  \bibfield  {author} {\bibinfo {author} {\bibfnamefont {M.~S.}\ \bibnamefont
  {Rudner}}, \bibinfo {author} {\bibfnamefont {N.~H.}\ \bibnamefont {Lindner}},
  \bibinfo {author} {\bibfnamefont {E.}~\bibnamefont {Berg}}, \ and\ \bibinfo
  {author} {\bibfnamefont {M.}~\bibnamefont {Levin}},\ }\href@noop {}
  {\bibfield  {journal} {\bibinfo  {journal} {Phys. Rev. X}\ }\textbf {\bibinfo
  {volume} {3}},\ \bibinfo {pages} {031005} (\bibinfo {year}
  {2013})}\BibitemShut {NoStop}%
\bibitem [{\citenamefont {Titum}\ \emph {et~al.}(2016)\citenamefont {Titum},
  \citenamefont {Berg}, \citenamefont {Rudner}, \citenamefont {Refael},\ and\
  \citenamefont {Lindner}}]{titumphysical}%
  \BibitemOpen
  \bibfield  {author} {\bibinfo {author} {\bibfnamefont {P.}~\bibnamefont
  {Titum}}, \bibinfo {author} {\bibfnamefont {E.}~\bibnamefont {Berg}},
  \bibinfo {author} {\bibfnamefont {M.~S.}\ \bibnamefont {Rudner}}, \bibinfo
  {author} {\bibfnamefont {G.}~\bibnamefont {Refael}}, \ and\ \bibinfo {author}
  {\bibfnamefont {N.~H.}\ \bibnamefont {Lindner}},\ }\href@noop {} {\bibfield
  {journal} {\bibinfo  {journal} {Phys. Rev. X}\ }\textbf {\bibinfo {volume}
  {6}},\ \bibinfo {pages} {021013} (\bibinfo {year} {2016})}\BibitemShut
  {NoStop}%
\bibitem [{\citenamefont {Leykam}\ \emph {et~al.}(2016)\citenamefont {Leykam},
  \citenamefont {Rechtsman},\ and\ \citenamefont
  {Chong}}]{leykam2016anomalous}%
  \BibitemOpen
  \bibfield  {author} {\bibinfo {author} {\bibfnamefont {D.}~\bibnamefont
  {Leykam}}, \bibinfo {author} {\bibfnamefont {M.~C.}\ \bibnamefont
  {Rechtsman}}, \ and\ \bibinfo {author} {\bibfnamefont {Y.~D.}\ \bibnamefont
  {Chong}},\ }\href@noop {} {\bibfield  {journal} {\bibinfo  {journal} {Phys.
  Rev. Lett.}\ }\textbf {\bibinfo {volume} {117}},\ \bibinfo {pages} {013902}
  (\bibinfo {year} {2016})}\BibitemShut {NoStop}%
\bibitem [{\citenamefont {Maczewsky}\ \emph {et~al.}(2017)\citenamefont
  {Maczewsky}, \citenamefont {Zeuner}, \citenamefont {Nolte},\ and\
  \citenamefont {Szameit}}]{maczewsky2017observation}%
  \BibitemOpen
  \bibfield  {author} {\bibinfo {author} {\bibfnamefont {L.~J.}\ \bibnamefont
  {Maczewsky}}, \bibinfo {author} {\bibfnamefont {J.~M.}\ \bibnamefont
  {Zeuner}}, \bibinfo {author} {\bibfnamefont {S.}~\bibnamefont {Nolte}}, \
  and\ \bibinfo {author} {\bibfnamefont {A.}~\bibnamefont {Szameit}},\
  }\href@noop {} {\bibfield  {journal} {\bibinfo  {journal} {Nature
  communications}\ }\textbf {\bibinfo {volume} {8}},\ \bibinfo {pages} {1}
  (\bibinfo {year} {2017})}\BibitemShut {NoStop}%
\bibitem [{\citenamefont {Kundu}\ \emph {et~al.}(2020)\citenamefont {Kundu},
  \citenamefont {Rudner}, \citenamefont {Berg},\ and\ \citenamefont
  {Lindner}}]{kundu2020quantized}%
  \BibitemOpen
  \bibfield  {author} {\bibinfo {author} {\bibfnamefont {A.}~\bibnamefont
  {Kundu}}, \bibinfo {author} {\bibfnamefont {M.}~\bibnamefont {Rudner}},
  \bibinfo {author} {\bibfnamefont {E.}~\bibnamefont {Berg}}, \ and\ \bibinfo
  {author} {\bibfnamefont {N.~H.}\ \bibnamefont {Lindner}},\ }\href@noop {}
  {\bibfield  {journal} {\bibinfo  {journal} {Phys. Rev. B}\ }\textbf {\bibinfo
  {volume} {101}},\ \bibinfo {pages} {041403} (\bibinfo {year}
  {2020})}\BibitemShut {NoStop}%
\bibitem [{\citenamefont {Nathan}\ \emph
  {et~al.}(2019{\natexlab{a}})\citenamefont {Nathan}, \citenamefont {Abanin},
  \citenamefont {Berg}, \citenamefont {Lindner},\ and\ \citenamefont
  {Rudner}}]{nathan2019anomalous}%
  \BibitemOpen
  \bibfield  {author} {\bibinfo {author} {\bibfnamefont {F.}~\bibnamefont
  {Nathan}}, \bibinfo {author} {\bibfnamefont {D.}~\bibnamefont {Abanin}},
  \bibinfo {author} {\bibfnamefont {E.}~\bibnamefont {Berg}}, \bibinfo {author}
  {\bibfnamefont {N.~H.}\ \bibnamefont {Lindner}}, \ and\ \bibinfo {author}
  {\bibfnamefont {M.~S.}\ \bibnamefont {Rudner}},\ }\href@noop {} {\bibfield
  {journal} {\bibinfo  {journal} {Phys. Rev. B}\ }\textbf {\bibinfo {volume}
  {99}},\ \bibinfo {pages} {195133} (\bibinfo {year}
  {2019}{\natexlab{a}})}\BibitemShut {NoStop}%
\bibitem [{\citenamefont {Duschatko}\ \emph {et~al.}(2018)\citenamefont
  {Duschatko}, \citenamefont {Dumitrescu},\ and\ \citenamefont
  {Potter}}]{PhysRevB.98.054309}%
  \BibitemOpen
  \bibfield  {author} {\bibinfo {author} {\bibfnamefont {B.~R.}\ \bibnamefont
  {Duschatko}}, \bibinfo {author} {\bibfnamefont {P.~T.}\ \bibnamefont
  {Dumitrescu}}, \ and\ \bibinfo {author} {\bibfnamefont {A.~C.}\ \bibnamefont
  {Potter}},\ }\href@noop {} {\bibfield  {journal} {\bibinfo  {journal} {Phys.
  Rev. B}\ }\textbf {\bibinfo {volume} {98}},\ \bibinfo {pages} {054309}
  (\bibinfo {year} {2018})}\BibitemShut {NoStop}%
\bibitem [{\citenamefont {Fidkowski}\ \emph {et~al.}(2019)\citenamefont
  {Fidkowski}, \citenamefont {Po}, \citenamefont {Potter},\ and\ \citenamefont
  {Vishwanath}}]{fidkowski2019interacting}%
  \BibitemOpen
  \bibfield  {author} {\bibinfo {author} {\bibfnamefont {L.}~\bibnamefont
  {Fidkowski}}, \bibinfo {author} {\bibfnamefont {H.~C.}\ \bibnamefont {Po}},
  \bibinfo {author} {\bibfnamefont {A.~C.}\ \bibnamefont {Potter}}, \ and\
  \bibinfo {author} {\bibfnamefont {A.}~\bibnamefont {Vishwanath}},\
  }\href@noop {} {\bibfield  {journal} {\bibinfo  {journal} {Phys. Rev. B}\
  }\textbf {\bibinfo {volume} {99}},\ \bibinfo {pages} {085115} (\bibinfo
  {year} {2019})}\BibitemShut {NoStop}%
\bibitem [{\citenamefont {Sieberer}\ \emph {et~al.}(2018)\citenamefont
  {Sieberer}, \citenamefont {Rieder}, \citenamefont {Fischer},\ and\
  \citenamefont {Fulga}}]{sieberer2018statistical}%
  \BibitemOpen
  \bibfield  {author} {\bibinfo {author} {\bibfnamefont {L.~M.}\ \bibnamefont
  {Sieberer}}, \bibinfo {author} {\bibfnamefont {M.-T.}\ \bibnamefont
  {Rieder}}, \bibinfo {author} {\bibfnamefont {M.~H.}\ \bibnamefont {Fischer}},
  \ and\ \bibinfo {author} {\bibfnamefont {I.~C.}\ \bibnamefont {Fulga}},\
  }\href@noop {} {\bibfield  {journal} {\bibinfo  {journal} {Phys. Rev. B}\
  }\textbf {\bibinfo {volume} {98}},\ \bibinfo {pages} {214301} (\bibinfo
  {year} {2018})}\BibitemShut {NoStop}%
\bibitem [{\citenamefont {Rieder}\ \emph {et~al.}(2018)\citenamefont {Rieder},
  \citenamefont {Sieberer}, \citenamefont {Fischer},\ and\ \citenamefont
  {Fulga}}]{rieder2018localization}%
  \BibitemOpen
  \bibfield  {author} {\bibinfo {author} {\bibfnamefont {M.-T.}\ \bibnamefont
  {Rieder}}, \bibinfo {author} {\bibfnamefont {L.~M.}\ \bibnamefont
  {Sieberer}}, \bibinfo {author} {\bibfnamefont {M.~H.}\ \bibnamefont
  {Fischer}}, \ and\ \bibinfo {author} {\bibfnamefont {I.~C.}\ \bibnamefont
  {Fulga}},\ }\href@noop {} {\bibfield  {journal} {\bibinfo  {journal} {Phys.
  Rev. Lett.}\ }\textbf {\bibinfo {volume} {120}},\ \bibinfo {pages} {216801}
  (\bibinfo {year} {2018})}\BibitemShut {NoStop}%
\bibitem [{Note1()}]{Note1}%
  \BibitemOpen
  \bibinfo {note} {Note that one can equivalently keep each step fixed at $T/5$
  and rescale the Hamiltonian by a factor $1+\delta _\ell $.}\BibitemShut
  {Stop}%
\bibitem [{Sup()}]{Supplement_PRL_Noisy_AFAI}%
  \BibitemOpen
  \href@noop {} {}\bibinfo {note} {Please see the supplementary online material
  (SOM) for a detailed discussion of the noise-averaged Floquet superoperator
  in the absence of spatial disorder.}\BibitemShut {Stop}%
\bibitem [{\citenamefont {Kolodrubetz}\ \emph {et~al.}(2018)\citenamefont
  {Kolodrubetz}, \citenamefont {Nathan}, \citenamefont {Gazit}, \citenamefont
  {Morimoto},\ and\ \citenamefont {Moore}}]{kolodrubetz2018topological}%
  \BibitemOpen
  \bibfield  {author} {\bibinfo {author} {\bibfnamefont {M.~H.}\ \bibnamefont
  {Kolodrubetz}}, \bibinfo {author} {\bibfnamefont {F.}~\bibnamefont {Nathan}},
  \bibinfo {author} {\bibfnamefont {S.}~\bibnamefont {Gazit}}, \bibinfo
  {author} {\bibfnamefont {T.}~\bibnamefont {Morimoto}}, \ and\ \bibinfo
  {author} {\bibfnamefont {J.~E.}\ \bibnamefont {Moore}},\ }\href@noop {}
  {\bibfield  {journal} {\bibinfo  {journal} {Phys. Rev. Lett.}\ }\textbf
  {\bibinfo {volume} {120}},\ \bibinfo {pages} {150601} (\bibinfo {year}
  {2018})}\BibitemShut {NoStop}%
\bibitem [{\citenamefont {Oganesyan}\ and\ \citenamefont
  {Huse}(2007)}]{PhysRevB.75.155111}%
  \BibitemOpen
  \bibfield  {author} {\bibinfo {author} {\bibfnamefont {V.}~\bibnamefont
  {Oganesyan}}\ and\ \bibinfo {author} {\bibfnamefont {D.~A.}\ \bibnamefont
  {Huse}},\ }\href@noop {} {\bibfield  {journal} {\bibinfo  {journal} {Phys.
  Rev. b}\ }\textbf {\bibinfo {volume} {75}},\ \bibinfo {pages} {155111}
  (\bibinfo {year} {2007})}\BibitemShut {NoStop}%
\bibitem [{Note2()}]{Note2}%
  \BibitemOpen
  \bibinfo {note} {Note that, for both $W_T=0$ and finite $W_T$, the plateau
  value $Q^\ast $ is not a sensitive indicator of the phase transition,
  requiring inaccesibly large system size in order to see a sharp
  transition}\BibitemShut {NoStop}%
\bibitem [{\citenamefont {Lee}\ \emph {et~al.}(2019)\citenamefont {Lee},
  \citenamefont {Ahn}, \citenamefont {Zhou},\ and\ \citenamefont
  {Vishwanath}}]{Lee2019_1}%
  \BibitemOpen
  \bibfield  {author} {\bibinfo {author} {\bibfnamefont {J.~Y.}\ \bibnamefont
  {Lee}}, \bibinfo {author} {\bibfnamefont {J.}~\bibnamefont {Ahn}}, \bibinfo
  {author} {\bibfnamefont {H.}~\bibnamefont {Zhou}}, \ and\ \bibinfo {author}
  {\bibfnamefont {A.}~\bibnamefont {Vishwanath}},\ }\href {\doibase
  10.1103/PhysRevLett.123.206404} {\bibfield  {journal} {\bibinfo  {journal}
  {Phys. Rev. Lett.}\ }\textbf {\bibinfo {volume} {123}},\ \bibinfo {pages}
  {206404} (\bibinfo {year} {2019})}\BibitemShut {NoStop}%
\bibitem [{\citenamefont {Friedman}\ \emph {et~al.}(2020)\citenamefont
  {Friedman}, \citenamefont {Ware}, \citenamefont {Vasseur},\ and\
  \citenamefont {Potter}}]{Friedman2020a}%
  \BibitemOpen
  \bibfield  {author} {\bibinfo {author} {\bibfnamefont {A.~J.}\ \bibnamefont
  {Friedman}}, \bibinfo {author} {\bibfnamefont {B.}~\bibnamefont {Ware}},
  \bibinfo {author} {\bibfnamefont {R.}~\bibnamefont {Vasseur}}, \ and\
  \bibinfo {author} {\bibfnamefont {A.~C.}\ \bibnamefont {Potter}},\
  }\href@noop {} {\bibfield  {journal} {\bibinfo  {journal} {arXiv preprint
  arXiv:2009.03314}\ } (\bibinfo {year} {2020})}\BibitemShut {NoStop}%
\bibitem [{\citenamefont {Else}\ \emph
  {et~al.}(2020{\natexlab{b}})\citenamefont {Else}, \citenamefont {Ho},\ and\
  \citenamefont {Dumitrescu}}]{Else2020b}%
  \BibitemOpen
  \bibfield  {author} {\bibinfo {author} {\bibfnamefont {D.~V.}\ \bibnamefont
  {Else}}, \bibinfo {author} {\bibfnamefont {W.~W.}\ \bibnamefont {Ho}}, \ and\
  \bibinfo {author} {\bibfnamefont {P.~T.}\ \bibnamefont {Dumitrescu}},\
  }\href@noop {} {\bibfield  {journal} {\bibinfo  {journal} {Phys. Rev. X}\
  }\textbf {\bibinfo {volume} {10}},\ \bibinfo {pages} {021032} (\bibinfo
  {year} {2020}{\natexlab{b}})}\BibitemShut {NoStop}%
\bibitem [{\citenamefont {Nandy}\ \emph {et~al.}(2017)\citenamefont {Nandy},
  \citenamefont {Sen},\ and\ \citenamefont {Sen}}]{Nandy2017}%
  \BibitemOpen
  \bibfield  {author} {\bibinfo {author} {\bibfnamefont {S.}~\bibnamefont
  {Nandy}}, \bibinfo {author} {\bibfnamefont {A.}~\bibnamefont {Sen}}, \ and\
  \bibinfo {author} {\bibfnamefont {D.}~\bibnamefont {Sen}},\ }\href@noop {}
  {\bibfield  {journal} {\bibinfo  {journal} {Phys. Rev. X}\ }\textbf {\bibinfo
  {volume} {7}},\ \bibinfo {pages} {031034} (\bibinfo {year}
  {2017})}\BibitemShut {NoStop}%
\bibitem [{\citenamefont {Nathan}\ \emph {et~al.}(2017)\citenamefont {Nathan},
  \citenamefont {Rudner}, \citenamefont {Lindner}, \citenamefont {Berg},\ and\
  \citenamefont {Refael}}]{nathan2017quantized}%
  \BibitemOpen
  \bibfield  {author} {\bibinfo {author} {\bibfnamefont {F.}~\bibnamefont
  {Nathan}}, \bibinfo {author} {\bibfnamefont {M.~S.}\ \bibnamefont {Rudner}},
  \bibinfo {author} {\bibfnamefont {N.~H.}\ \bibnamefont {Lindner}}, \bibinfo
  {author} {\bibfnamefont {E.}~\bibnamefont {Berg}}, \ and\ \bibinfo {author}
  {\bibfnamefont {G.}~\bibnamefont {Refael}},\ }\href@noop {} {\bibfield
  {journal} {\bibinfo  {journal} {Phys. Rev. Lett.}\ }\textbf {\bibinfo
  {volume} {119}},\ \bibinfo {pages} {186801} (\bibinfo {year}
  {2017})}\BibitemShut {NoStop}%
\bibitem [{\citenamefont {Nathan}\ \emph
  {et~al.}(2019{\natexlab{b}})\citenamefont {Nathan}, \citenamefont {Abanin},
  \citenamefont {Lindner}, \citenamefont {Berg},\ and\ \citenamefont
  {Rudner}}]{nathan2019hierarchy}%
  \BibitemOpen
  \bibfield  {author} {\bibinfo {author} {\bibfnamefont {F.}~\bibnamefont
  {Nathan}}, \bibinfo {author} {\bibfnamefont {D.~A.}\ \bibnamefont {Abanin}},
  \bibinfo {author} {\bibfnamefont {N.~H.}\ \bibnamefont {Lindner}}, \bibinfo
  {author} {\bibfnamefont {E.}~\bibnamefont {Berg}}, \ and\ \bibinfo {author}
  {\bibfnamefont {M.~S.}\ \bibnamefont {Rudner}},\ }\href@noop {} {\bibfield
  {journal} {\bibinfo  {journal} {arXiv preprint arXiv:1907.12228}\ } (\bibinfo
  {year} {2019}{\natexlab{b}})}\BibitemShut {NoStop}%
\bibitem [{\citenamefont {Jiang}\ \emph {et~al.}(2011)\citenamefont {Jiang},
  \citenamefont {Kitagawa}, \citenamefont {Alicea}, \citenamefont {Akhmerov},
  \citenamefont {Pekker}, \citenamefont {Refael}, \citenamefont {Cirac},
  \citenamefont {Demler}, \citenamefont {Lukin},\ and\ \citenamefont
  {Zoller}}]{PhysRevLett.106.220402}%
  \BibitemOpen
  \bibfield  {author} {\bibinfo {author} {\bibfnamefont {L.}~\bibnamefont
  {Jiang}}, \bibinfo {author} {\bibfnamefont {T.}~\bibnamefont {Kitagawa}},
  \bibinfo {author} {\bibfnamefont {J.}~\bibnamefont {Alicea}}, \bibinfo
  {author} {\bibfnamefont {A.~R.}\ \bibnamefont {Akhmerov}}, \bibinfo {author}
  {\bibfnamefont {D.}~\bibnamefont {Pekker}}, \bibinfo {author} {\bibfnamefont
  {G.}~\bibnamefont {Refael}}, \bibinfo {author} {\bibfnamefont {J.~I.}\
  \bibnamefont {Cirac}}, \bibinfo {author} {\bibfnamefont {E.}~\bibnamefont
  {Demler}}, \bibinfo {author} {\bibfnamefont {M.~D.}\ \bibnamefont {Lukin}}, \
  and\ \bibinfo {author} {\bibfnamefont {P.}~\bibnamefont {Zoller}},\
  }\href@noop {} {\bibfield  {journal} {\bibinfo  {journal} {Phys. Rev. Lett.}\
  }\textbf {\bibinfo {volume} {106}},\ \bibinfo {pages} {220402} (\bibinfo
  {year} {2011})}\BibitemShut {NoStop}%
\bibitem [{\citenamefont {Bauer}\ \emph {et~al.}(2019)\citenamefont {Bauer},
  \citenamefont {Pereg-Barnea}, \citenamefont {Karzig}, \citenamefont {Rieder},
  \citenamefont {Refael}, \citenamefont {Berg},\ and\ \citenamefont
  {Oreg}}]{PhysRevB.100.041102}%
  \BibitemOpen
  \bibfield  {author} {\bibinfo {author} {\bibfnamefont {B.}~\bibnamefont
  {Bauer}}, \bibinfo {author} {\bibfnamefont {T.}~\bibnamefont {Pereg-Barnea}},
  \bibinfo {author} {\bibfnamefont {T.}~\bibnamefont {Karzig}}, \bibinfo
  {author} {\bibfnamefont {M.-T.}\ \bibnamefont {Rieder}}, \bibinfo {author}
  {\bibfnamefont {G.}~\bibnamefont {Refael}}, \bibinfo {author} {\bibfnamefont
  {E.}~\bibnamefont {Berg}}, \ and\ \bibinfo {author} {\bibfnamefont
  {Y.}~\bibnamefont {Oreg}},\ }\href@noop {} {\bibfield  {journal} {\bibinfo
  {journal} {Phys. Rev. B}\ }\textbf {\bibinfo {volume} {100}},\ \bibinfo
  {pages} {041102} (\bibinfo {year} {2019})}\BibitemShut {NoStop}%
\end{thebibliography}%


%


%


\begin{thebibliography}{5}%
\makeatletter
\providecommand \@ifxundefined [1]{%
 \@ifx{#1\undefined}
}%
\providecommand \@ifnum [1]{%
 \ifnum #1\expandafter \@firstoftwo
 \else \expandafter \@secondoftwo
 \fi
}%
\providecommand \@ifx [1]{%
 \ifx #1\expandafter \@firstoftwo
 \else \expandafter \@secondoftwo
 \fi
}%
\providecommand \natexlab [1]{#1}%
\providecommand \enquote  [1]{``#1''}%
\providecommand \bibnamefont  [1]{#1}%
\providecommand \bibfnamefont [1]{#1}%
\providecommand \citenamefont [1]{#1}%
\providecommand \href@noop [0]{\@secondoftwo}%
\providecommand \href [0]{\begingroup \@sanitize@url \@href}%
\providecommand \@href[1]{\@@startlink{#1}\@@href}%
\providecommand \@@href[1]{\endgroup#1\@@endlink}%
\providecommand \@sanitize@url [0]{\catcode `\\12\catcode `\$12\catcode
  `\&12\catcode `\#12\catcode `\^12\catcode `\_12\catcode `\%12\relax}%
\providecommand \@@startlink[1]{}%
\providecommand \@@endlink[0]{}%
\providecommand \url  [0]{\begingroup\@sanitize@url \@url }%
\providecommand \@url [1]{\endgroup\@href {#1}{\urlprefix }}%
\providecommand \urlprefix  [0]{URL }%
\providecommand \Eprint [0]{\href }%
\providecommand \doibase [0]{http://dx.doi.org/}%
\providecommand \selectlanguage [0]{\@gobble}%
\providecommand \bibinfo  [0]{\@secondoftwo}%
\providecommand \bibfield  [0]{\@secondoftwo}%
\providecommand \translation [1]{[#1]}%
\providecommand \BibitemOpen [0]{}%
\providecommand \bibitemStop [0]{}%
\providecommand \bibitemNoStop [0]{.\EOS\space}%
\providecommand \EOS [0]{\spacefactor3000\relax}%
\providecommand \BibitemShut  [1]{\csname bibitem#1\endcsname}%
\let\auto@bib@innerbib\@empty
\bibitem [{\citenamefont {Sieberer}\ \emph {et~al.}(2018)\citenamefont
  {Sieberer}, \citenamefont {Rieder}, \citenamefont {Fischer},\ and\
  \citenamefont {Fulga}}]{Sieberer2018}%
  \BibitemOpen
  \bibfield  {author} {\bibinfo {author} {\bibfnamefont {L.~M.}\ \bibnamefont
  {Sieberer}}, \bibinfo {author} {\bibfnamefont {M.-T.}\ \bibnamefont
  {Rieder}}, \bibinfo {author} {\bibfnamefont {M.~H.}\ \bibnamefont {Fischer}},
  \ and\ \bibinfo {author} {\bibfnamefont {I.~C.}\ \bibnamefont {Fulga}},\
  }\bibfield  {title} {\enquote {\bibinfo {title} {{Statistical periodicity in
  driven quantum systems: General formalism and application to noisy Floquet
  topological chains}},}\ }\href {\doibase 10.1103/PhysRevB.98.214301}
  {\bibfield  {journal} {\bibinfo  {journal} {Phys. Rev. B}\ }\textbf {\bibinfo
  {volume} {98}},\ \bibinfo {pages} {214301} (\bibinfo {year}
  {2018})}\BibitemShut {NoStop}%
\bibitem [{\citenamefont {Titum}\ \emph {et~al.}(2016)\citenamefont {Titum},
  \citenamefont {Berg}, \citenamefont {Rudner}, \citenamefont {Refael},\ and\
  \citenamefont {Lindner}}]{Titum2016}%
  \BibitemOpen
  \bibfield  {author} {\bibinfo {author} {\bibfnamefont {P.}~\bibnamefont
  {Titum}}, \bibinfo {author} {\bibfnamefont {E.}~\bibnamefont {Berg}},
  \bibinfo {author} {\bibfnamefont {M.~S.}\ \bibnamefont {Rudner}}, \bibinfo
  {author} {\bibfnamefont {G.}~\bibnamefont {Refael}}, \ and\ \bibinfo {author}
  {\bibfnamefont {N.~H.}\ \bibnamefont {Lindner}},\ }\bibfield  {title}
  {\enquote {\bibinfo {title} {{Anomalous Floquet-Anderson Insulator as a
  Nonadiabatic Quantized Charge Pump}},}\ }\href {\doibase
  10.1103/PhysRevX.6.021013} {\bibfield  {journal} {\bibinfo  {journal} {Phys.
  Rev. X}\ }\textbf {\bibinfo {volume} {6}},\ \bibinfo {pages} {021013}
  (\bibinfo {year} {2016})}\BibitemShut {NoStop}%
\bibitem [{\citenamefont {Landi}\ \emph {et~al.}(2014)\citenamefont {Landi},
  \citenamefont {Novais}, \citenamefont {de~Oliveira},\ and\ \citenamefont
  {Karevski}}]{Landi2014}%
  \BibitemOpen
  \bibfield  {author} {\bibinfo {author} {\bibfnamefont {G.~T.}\ \bibnamefont
  {Landi}}, \bibinfo {author} {\bibfnamefont {E.}~\bibnamefont {Novais}},
  \bibinfo {author} {\bibfnamefont {M.~J.}\ \bibnamefont {de~Oliveira}}, \ and\
  \bibinfo {author} {\bibfnamefont {D.}~\bibnamefont {Karevski}},\ }\bibfield
  {title} {\enquote {\bibinfo {title} {{Flux rectification in the quantum $XXZ$
  chain}},}\ }\href {\doibase 10.1103/PhysRevE.90.042142} {\bibfield  {journal}
  {\bibinfo  {journal} {Phys. Rev. E}\ }\textbf {\bibinfo {volume} {90}},\
  \bibinfo {pages} {042142} (\bibinfo {year} {2014})}\BibitemShut {NoStop}%
\bibitem [{\citenamefont {Rudner}\ \emph {et~al.}(2013)\citenamefont {Rudner},
  \citenamefont {Lindner}, \citenamefont {Berg},\ and\ \citenamefont
  {Levin}}]{Rudner2013}%
  \BibitemOpen
  \bibfield  {author} {\bibinfo {author} {\bibfnamefont {M.~S.}\ \bibnamefont
  {Rudner}}, \bibinfo {author} {\bibfnamefont {N.~H.}\ \bibnamefont {Lindner}},
  \bibinfo {author} {\bibfnamefont {E.}~\bibnamefont {Berg}}, \ and\ \bibinfo
  {author} {\bibfnamefont {M.}~\bibnamefont {Levin}},\ }\bibfield  {title}
  {\enquote {\bibinfo {title} {{Anomalous Edge States and the Bulk-Edge
  Correspondence for Periodically Driven Two-Dimensional Systems}},}\ }\href
  {\doibase 10.1103/PhysRevX.3.031005} {\bibfield  {journal} {\bibinfo
  {journal} {Phys. Rev. X}\ }\textbf {\bibinfo {volume} {3}},\ \bibinfo {pages}
  {031005} (\bibinfo {year} {2013})}\BibitemShut {NoStop}%
\bibitem [{\citenamefont {Rieder}\ \emph {et~al.}(2018)\citenamefont {Rieder},
  \citenamefont {Sieberer}, \citenamefont {Fischer},\ and\ \citenamefont
  {Fulga}}]{Rieder2017}%
  \BibitemOpen
  \bibfield  {author} {\bibinfo {author} {\bibfnamefont {M.-T.}\ \bibnamefont
  {Rieder}}, \bibinfo {author} {\bibfnamefont {L.~M.}\ \bibnamefont
  {Sieberer}}, \bibinfo {author} {\bibfnamefont {M.~H.}\ \bibnamefont
  {Fischer}}, \ and\ \bibinfo {author} {\bibfnamefont {I.~C.}\ \bibnamefont
  {Fulga}},\ }\bibfield  {title} {\enquote {\bibinfo {title} {{Localization
  Counteracts Decoherence in Noisy Floquet Topological Chains}},}\ }\href
  {\doibase 10.1103/PhysRevLett.120.216801} {\bibfield  {journal} {\bibinfo
  {journal} {Phys. Rev. Lett.}\ }\textbf {\bibinfo {volume} {120}},\ \bibinfo
  {pages} {216801} (\bibinfo {year} {2018})}\BibitemShut {NoStop}%
\end{thebibliography}%


%


%


\end{document}


\title{Supplemental Material: Quantized Floquet topology with temporal noise}

\author{Christopher I. Timms}

\affiliation{Department of Physics, University of Texas at Dallas,
  Richardson, TX, USA}

\author{Lukas M. Sieberer}

\affiliation{Institute for Theoretical Physics, University of Innsbruck, 6020
  Innsbruck, Austria}

\author{Michael H. Kolodrubetz}

\affiliation{Department of Physics, University of Texas at Dallas,
  Richardson, TX, USA}

\maketitle

In this supplement, we discuss how the phase transition in the clean system
with $W=0$ can be captured directly from the Floquet superoperator approach
introduced in Ref.~\cite{Sieberer2018}. At the phase transition, a gap in the complex spectrum of the noise-averaged Floquet superoperator closes, and, concomitantly, a topological invariant defined in terms of the Floquet superoperator loses its quantization. The techniques we introduce in the following will allow numerical solutions in the presence of disorder, although we do not treat this more complicated case in the present work.

\section{Floquet superoperator}
\label{sec:floq-super}

We start by considering the AFAI Hamiltonian~\cite{Titum2016} in
the clean limit in the single-particle basis formed by the states $\ket{\mathbf{k}} \otimes \ket{s}$
with quasimomentum $\mathbf{k} = \left( k_x, k_y \right)^{\transpose}$ and
sublattice index $s = A, B$. Under noise-averaged dynamics as specified below,
the time-evolved state of a single particle becomes mixed, and can be described
by a density matrix of the form
\begin{equation}
  \rho = \sum_{\mathbf{k}_1, \mathbf{k}_2} \ket{\mathbf{k}_1} \bra{\mathbf{k}_2} \otimes
  \rho_{\mathbf{k}_1, \mathbf{k}_2}, 
\end{equation}
where $\rho_{\mathbf{k}_1, \mathbf{k}_2}$ is a $2 \times 2$ matrix in sublattice
space.

The dynamics of the density matrix is governed by the von Neumann equation. For momentum modes
$\mathbf{k}_1$ and $\mathbf{k}_2$, this is
\begin{equation}
  \label{eq:von-Neumann}
  \frac{\mathrm{d} \rho_{\mathbf{k}_1, \mathbf{k}_2}}{\mathrm{d} t} = - \imag \left(
    H_{\mathbf{k}_1} \rho_{\mathbf{k}_1, \mathbf{k}_2} - \rho_{\mathbf{k}_1,
      \mathbf{k}_2} H_{\mathbf{k}_2} \right).
\end{equation}
We can vectorize the density matrix by stacking columns of
$\rho_{\mathbf{k}_1, \mathbf{k}_2}$ (see, for example, Ref.~\cite{Landi2014}) to
get
\begin{equation}
  \label{eq:rho-vectorized}
  \kket{\rho_{\mathbf{k}_1, \mathbf{k}_2}} = \left( \rho^{A, A}_{\mathbf{k}_1,
      \mathbf{k}_2}, \rho^{B, A}_{\mathbf{k}_1, \mathbf{k}_2}, \rho^{A,
      B}_{\mathbf{k}_1, \mathbf{k}_2}, \rho^{B, B}_{\mathbf{k}_1, \mathbf{k}_2} \right)^{\transpose}.
\end{equation}
The basic relation to rewrite the von Neumann equation~\eqref{eq:von-Neumann}
for the density matrix in vectorized form reads
\begin{equation}
  \kket{A \rho B} = B^{\transpose} \otimes A \kket{\rho}.
\end{equation}
By using this relation, Eq.~\eqref{eq:von-Neumann} can be recast as 
\begin{equation}
  \label{eq:von-Neumann-vectorized}
  \frac{\mathrm{d}}{\mathrm{d} t} \kket{\rho_{\mathbf{k}_1, \mathbf{k}_2}} = - \imag
  \mathcal{H}_{\mathbf{k}_1, \mathbf{k}_2} \kket{\rho_{\mathbf{k}_1, \mathbf{k}_2}},
\end{equation}
where
\begin{equation}
  \mathcal{H}_{\mathbf{k}_1, \mathbf{k}_2} = \id \otimes H_{\mathbf{k}_1} -
  H_{\mathbf{k}_2}^{\transpose} \otimes \id.
\end{equation}

The AFAI without disorder is described by a periodically time-dependent Bloch
Hamiltonian $H_{\mathbf{k}}(t) = H_{\mathbf{k}, m}$ for
$\left( m - 1 \right) T/M \leq t < m T/M$, where $m = 1, \dotsc, M$, and
$M = 5$ is the number of steps of the Floquet drive. In particular, for $m = 1, \dotsc, 4$, we set~\cite{Rudner2013}
\begin{equation}
  H_{\mathbf{k}, m} = - J \left( \e^{\imag \mathbf{b}_m \cdot \mathbf{k}}
    \sigma_+ + \e^{-\imag \mathbf{b}_m \cdot \mathbf{k}} \sigma_- \right),  
\end{equation}
where $\mathbf{b}_1 = - \mathbf{b}_3 = \left( 1, 0 \right)$ and
$\mathbf{b}_2 = - \mathbf{b}_4 = \left( 0, 1 \right)$ (we set the lattice constant to $a = 1$), and
\begin{equation}
  H_{\mathbf{k}, 5} = \Delta \sigma_z.
\end{equation}

As in the main text, we consider the case of timing noise:
$\mathcal{H}_{\mathbf{k}, m}$ is applied for
$T_m = T \left( 1 + \delta_m \right) \! /M$, where the random time shifts
$\delta_m$ are sampled uniformly from a finite interval,
$\delta_m \in [- W_{\mathrm{T}}, W_{\mathrm{T}}]$. By integrating
Eq.~\eqref{eq:von-Neumann-vectorized} over one noisy sequence of steps
$1 \to 2 \to \dotsb \to M = 5$, we obtain
\begin{equation}
  \label{eq:noisy-U}
  \mathcal{U}_{\mathbf{k}_1, \mathbf{k}_2} = \e^{- \imag T_M \mathcal{H}_{\mathbf{k}_1, \mathbf{k}_2, M}} \dotsb
  \e^{- \imag T_1 \mathcal{H}_{\mathbf{k}_1, \mathbf{k}_2, 1}},
\end{equation}
where
\begin{equation}
  \mathcal{H}_{\mathbf{k}_1, \mathbf{k}_2, m} = \id \otimes H_{\mathbf{k}_1, m} -
  H_{\mathbf{k}_2, m}^{\transpose} \otimes \id.
\end{equation}
The average of Eq.~\eqref{eq:noisy-U} over noise realizations yields the Floquet
superoperator~\cite{Rieder2017, Sieberer2018}:
\begin{multline}
  \label{eq:F}
  \mathcal{F}_{\mathbf{k}_1, \mathbf{k}_2} =
  \overline{\mathcal{U}_{\mathbf{k}_1, \mathbf{k}_2}} \\ = \e^{-\imag T
    \mathcal{H}_{\mathbf{k}_1, \mathbf{k}_2, M}/M} \mathcal{E}_{\mathbf{k}_1,
    \mathbf{k}_2, M} \dotsb \e^{-\imag T \mathcal{H}_{\mathbf{k}_1,
      \mathbf{k}_2, 1}/M} \mathcal{E}_{\mathbf{k}, \mathbf{k}', 1},
\end{multline}
where
\begin{equation}
\label{eq:E}
  \begin{split}
    \mathcal{E}_{\mathbf{k}_1, \mathbf{k}_2, m} & = \overline{\e^{-\imag T \delta_m
        \mathcal{H}_{\mathbf{k}_1, \mathbf{k}_2, m}/M}} \\ & = \frac{1}{2 W_{\mathrm{T}}}
    \int_{-W_{\mathrm{T}}}^{W_{\mathrm{T}}} \mathrm{d} \delta_m \, \e^{-\imag T
      \delta_m \mathcal{H}_{\mathbf{k}_1, \mathbf{k}_2, m}/M} \\ & = \sinc(T W_{\mathrm{T}}
    \mathcal{H}_{\mathbf{k}_1, \mathbf{k}_2, m}/M),
  \end{split}
\end{equation}
and $\mathrm{sinc}(x) = \sin(x)/x$.

\section{Complex Floquet-Bloch bands}
\label{sec:compl-floq-bloch}

We proceed to determine the spectrum of the Floquet superoperator. In analogy to
the case of a conventional unitary Floquet operator such as
\begin{equation}
  U_{\mathbf{k}} = \e^{-\imag T H_{\mathbf{k}, M}/M} \dotsb \e^{-\imag T
    H_{\mathbf{k}, 1}/M},
\end{equation}
which has unimodular eigenvalues $\e^{- \imag T \varepsilon_{n, \mathbf{k}}}$
with Floquet-Bloch bands $\varepsilon_{n, \mathbf{k}}$, we write the
eigenequation of the Floquet superoperator in the form
\begin{equation}
  \label{eq:eigenequation-F}
  \mathcal{F}_{\mathbf{k}_1, \mathbf{k}_2} \kket{\rho_{n, \mathbf{k}_1, \mathbf{k}_2}}
  = \e^{-\imag T \lambda_{n, \mathbf{k}_1, \mathbf{k}_2}} \kket{\rho_{n,
      \mathbf{k}_1, \mathbf{k}_2}}.
\end{equation}
Here, $n = 1, 2, 3, 4$ labels the bands $\lambda_{n, \mathbf{k}_1,
  \mathbf{k}_2}$, which are, in general, functions of both $\mathbf{k}_1$ and
$\mathbf{k}_2$.

It is instructive to consider first the case $W_{\mathrm{T}} = 0$. Then, Eq.~\eqref{eq:E} reduces to $\mathcal{E}_{\mathbf{k}_1, \mathbf{k}_2, m} = \id$. Further, using the
relation
\begin{equation}
  \begin{split}
    \e^{- \imag T \mathcal{H}_{\mathbf{k}_1, \mathbf{k}_2, m}/M} & = \e^{- \imag
      T \left( \id \otimes H_{\mathbf{k}_1} \right)/M} \e^{\imag T \left(
        H_{\mathbf{k}_2}^{\transpose} \otimes \id \right)/M} \\ & =  \e^{\imag T
      H_{\mathbf{k}_2, m}^{\transpose}/M} \otimes \e^{- \imag T H_{\mathbf{k}_1,
        m}/M},
  \end{split}
\end{equation}
allows us to factorize the Floquet superoperator in Eq.~\eqref{eq:F} as
\begin{equation}
  \label{eq:F-unitary}
  \mathcal{F}_{\mathbf{k}_1, \mathbf{k}_2} = U_{\mathbf{k}_2}^{*} \otimes U_{\mathbf{k}_1}.
\end{equation}
The eigenvalues of $\mathcal{F}_{\mathbf{k}_1, \mathbf{k}_2}$ are thus given by
products of eigenvalues of $U_{\mathbf{k}}$,
\begin{equation}
  \e^{-\imag T \lambda_{n, \mathbf{k}_1, \mathbf{k}_2}} = \e^{-\imag T \left(
      \varepsilon_{n_1, \mathbf{k}_1} - \varepsilon_{n_2, \mathbf{k}_2}
    \right)},
\end{equation}
for all combinations of $n_1, n_2 = 1, 2$. Therefore, if $W_{\mathrm{T}} = 0$,
all $\lambda_{n, \mathbf{k}_1, \mathbf{k}_2}$ in Eq.~\eqref{eq:eigenequation-F}
are real, and the Floquet superoperator is unitary. If $W_T \neq 0$, then the
$\lambda_{n, \mathbf{k}_1, \mathbf{k}_2}$ are in general complex, indicating
that averaging over timing noise causes certain bulk states to decay in time.

\begin{figure}
  \centering
  \includegraphics{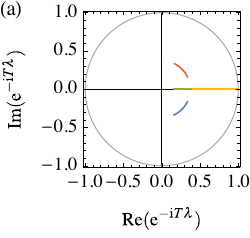}
  \includegraphics{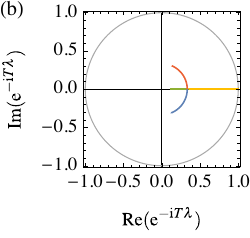}
  \includegraphics{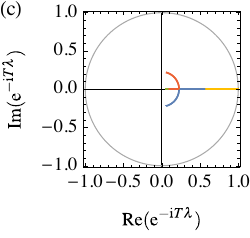}
  \includegraphics{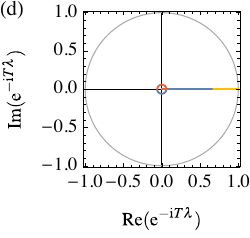}
  \caption{Spectrum of the Floquet superoperator. (a)
    $W_{\mathrm{T}} = 0.51 \Omega$. Two complex bands are shown as blue and red
    lines. For the bands which are shown as yellow and green lines, $\lambda$ is
    purely imaginary. (b) At the first critical value
    $W_{\mathrm{T}} = W_{\mathrm{T, c, 1}}$, the gap along the positive real
    axis between the complex bands closes. (c) For
    $W_{\mathrm{T}} = 0.61 \Omega > W_{\mathrm{T, c, 1}}$, portions of the
    formerly complex bands become purely imaginary. (d) At the second critical
    value $W_{\mathrm{T}} = W_{\mathrm{T, c, 2}}$, there is a closing of the gap
    along the negative real axis.}
  \label{fig:F-spectrum}
\end{figure}

The complex spectrum of the Floquet superoperator for $T_m = T/5$,
$J = 5 \Omega/4$, and $\Delta = 2 \Omega/5$, where $\Omega = 2 \pi/T$, is illustrated in
Fig.~\ref{fig:F-spectrum} for
various values of $W_{\mathrm{T}} \neq 0$. We can identify two critical values of
$W_{\mathrm{T}}$: For $W_{\mathrm{T}} = W_{\mathrm{T, c, 1}}$, which is shown in
Fig.~\ref{fig:F-spectrum}(b), the gap along the positive real axis between the
bands shown as blue and red lines closes. Further, for
$W_{\mathrm{T}} = W_{\mathrm{T, c, 2}}$, shown in Fig.~\ref{fig:F-spectrum}(d),
there is a closing of the gap along the negative real axis between the same two
bands. To elucidate the role which the various bands play in the dynamics of the
system, and to determine the critical values of $W_{\mathrm{T}}$, we proceed to
study the structure of the Floquet superoperator in more detail.

With the above choice of $T_m$, $J$, and $\Delta$, and for any value of
$W_{\mathrm{T}}$, the Floquet superoperator takes a rather simple form:
\begin{equation}
  \mathcal{F}_{\mathbf{k}_1, \mathbf{k}_2} =
  \begin{pmatrix}
    \mathcal{F}_{1, 1, \mathbf{k}} & 0 & 0 & \mathcal{F}_{1, 4, \mathbf{k}} \\
    0 & \mathcal{F}_{2, 2, \mathbf{K}} & \mathcal{F}_{2, 3, \mathbf{K}} & 0 \\
    0 & \mathcal{F}_{3, 2, \mathbf{K}} & \mathcal{F}_{3, 3, \mathbf{K}} & 0 \\
    \mathcal{F}_{4, 1, \mathbf{k}} & 0 & 0 & \mathcal{F}_{4, 4, \mathbf{k}}
  \end{pmatrix},
\end{equation}
where the respective matrix elements depend only on the sum or difference of
$\mathbf{k}_1$ and $\mathbf{k}_2$,
\begin{equation}
  \mathbf{K} = \frac{\mathbf{k}_1 + \mathbf{k}_2}{2}, \qquad \mathbf{k} =
  \mathbf{k}_1 - \mathbf{k}_2.
\end{equation}
With regard to its sublattice structure, the Floquet superoperator is
composed of two decoupled $2 \times 2$ blocks, each giving rise to two
bands. We denote the bands that are determined by the outer block (the first and fourth rows and columns of $\mathcal{F}_{\mathbf{k}_1, \mathbf{k}_2}$) by
$\lambda_{1, \mathbf{k}}$ and $\lambda_{2, \mathbf{k}}$, and those that are determined by the inner block by $\lambda_{3,
  \mathbf{K}}$ and $\lambda_{4, \mathbf{K}}$.

\begin{figure}
  \centering
  \includegraphics{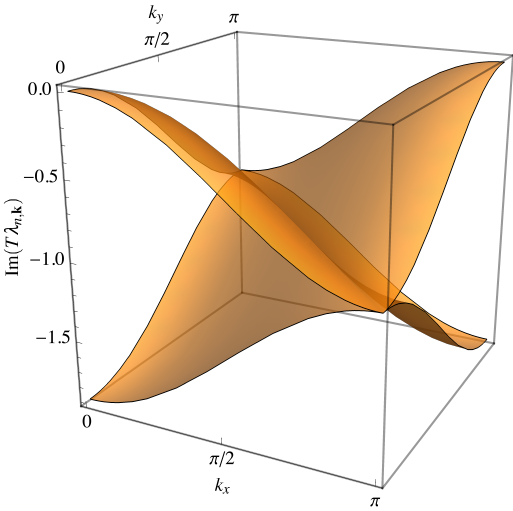}  
  \caption{Floquet-Bloch bands $\lambda_{n, \mathbf{k}}$ with $n = 1, 2$ for
    $W_{\mathrm{T}} = 0.51 \Omega$.}
  \label{fig:lambda-1-2}
\end{figure}

\begin{figure*}
  \centering
  \includegraphics{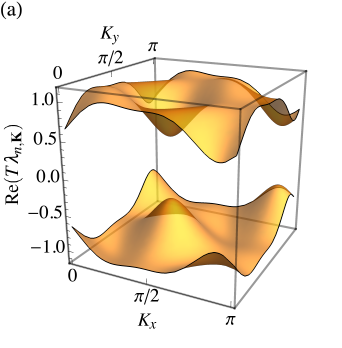}
  \includegraphics{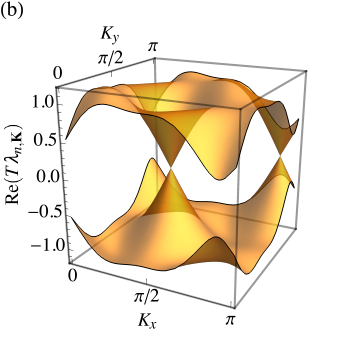} 
  \includegraphics{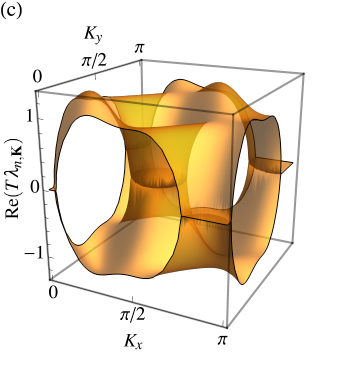} \\
  \includegraphics{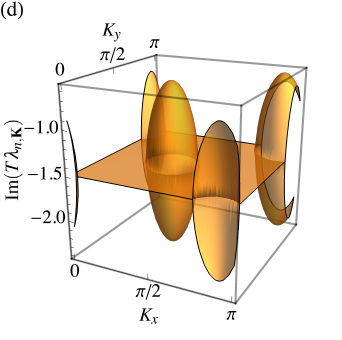}
  \includegraphics{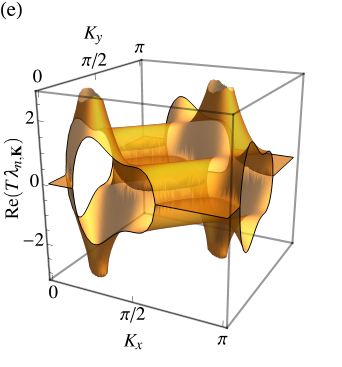}
  \includegraphics{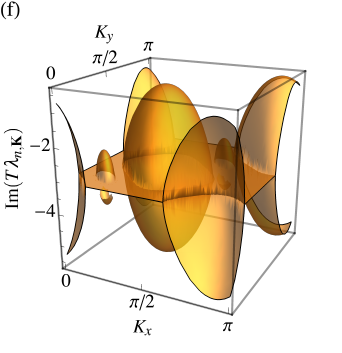}
  \caption{Floquet-Bloch bands $\lambda_{n, \mathbf{K}}$ with $n = 3, 4$. (a)
    For $W_{\mathrm{T}} = 0.51 \Omega < W_{\mathrm{T, c, 1}}$, both bands are
    complex with nonvanishing real and imaginary parts. We only show the real part. The constant imaginary part is given in Eq.~\eqref{eq:lambda-3-4-imag}. (b) At the first critical value
    $W_{\mathrm{T}} = W_{\mathrm{T,c,1}}$, the gap at
    $\Re(T \lambda_{n, \mathbf{K}}) = 0$ closes. For
    $W_{\mathrm{T}} = 0.61 \Omega > W_{\mathrm{T, c, 1}}$, in finite regions of
    the Brillouin zone, (c) the real parts vanish, and (d) the bands are purely
    imaginary with nonconstant imaginary part. The case $W_{\mathrm{T}} > W_{\mathrm{T, c, 2}}$, in which also
    the gap at $\Re(T \lambda_{n, \mathbf{K}}) = \pi$ is closed, is
    illustrated in (e) and (f).}
  \label{fig:lambda-3-4}
\end{figure*}

From the vectorized form of the density matrix in Eq.~\eqref{eq:rho-vectorized},
it follows that the dynamics of the elements of the density matrix that are
diagonal in sublattice space is governed by the outer block of the Floquet
superoperator. The bands $\lambda_{1, \mathbf{k}}$ and
$\lambda_{2, \mathbf{k}}$, which are encoded in this block and which are shown
in Fig.~\ref{fig:F-spectrum} as yellow and green lines, respectively, have
vanishing real parts. The imaginary parts are illustrated in
Fig.~\ref{fig:lambda-1-2}. In particular, the values $\lambda_{1, \mathbf{k}}$
and $\lambda_{2, \mathbf{k}}$ for $\mathbf{k} = \mathbf{0}$ determine the
dynamics of populations, i.e., of the elements of the density matrix which are
diagonal both in sublattice and momentum space. These values are
\begin{equation}
  \lambda_{1, \mathbf{0}} = 0, \qquad \lambda_{2, \mathbf{0}} = \imag \frac{4}{T}
  \ln(\sinc(\pi W_{\mathrm{T}})).
\end{equation}
The corresponding eigenvectors are, in their non-vectorized forms,
\begin{equation}
  \rho_{1, \mathbf{0}} = \frac{\id}{\sqrt{2}}, \qquad
  \rho_{2, \mathbf{0}} = \frac{1}{\sqrt{2}}
  \begin{pmatrix}
    1 & 0 \\ 0 & - 1
  \end{pmatrix}.
\end{equation}
Thus, the eigenvalue $\lambda_{1, \mathbf{0}} = 0$ corresponds to the fully mixed
steady state, while $\lambda_{2, \mathbf{0}}$ determines the rate at which any
initial imbalance of populations decays.

The central $2 \times 2$ block of $\mathcal{F}_{\mathbf{k}_1, \mathbf{k}_2}$
determines the dynamics of elements of the density matrix that are off-diagonal
in sublattice space. The corresponding bands $\lambda_{3, \mathbf{K}}$ and
$\lambda_{4, \mathbf{K}}$ are shown in Fig.~\ref{fig:lambda-3-4}. For small
values of $W_{\mathrm{T}}$, these bands have equal and constant imaginary parts,
\begin{equation}
\label{eq:lambda-3-4-imag}
  \Im(\lambda_{n, \mathbf{K}}) = \frac{\imag}{T} \ln \! \left( \sinc(8 \pi
    W_{\mathrm{T}}/25) \sinc(\pi W_{\mathrm{T}})^2 \right).
\end{equation}
To determine the critical disorder strengths $W_{\mathrm{T}, \mathrm{c}}$ and
the momenta
$\mathbf{K}_{\mathrm{c}} = \left( K_{\mathrm{c}, x}, K_{\mathrm{c}, y} \right)$
at which gap closings occur, we note that $\lambda_{n, \mathbf{K}}$ is symmetric
under the exchange of $K_x$ and $K_y$. Therefore, gap closings must occur as
pairs
$\mathbf{K}_{\mathrm{c}, 1} = \left( K_{\mathrm{c}, x}, K_{\mathrm{c}, y}
\right)$
and
$\mathbf{K}_{\mathrm{c}, 2} = \left( K_{\mathrm{c}, y}, K_{\mathrm{c}, x}
\right)$,
or on the line $K_{\mathrm{c}} = K_{\mathrm{c} x} = K_{\mathrm{c}, y}$.
Inspection of the band structures in Fig.~\ref{fig:lambda-3-4} indicates that
the latter is the case. Therefore, we consider the difference of eigenvalues of
the Floquet superoperator for $K = K_x = K_y$,
\begin{multline}
  \e^{-\imag T \lambda_{3, \mathbf{K}}} - \e^{-\imag T \lambda_{4, \mathbf{K}}}
  = - \imag \sinc \biggl( \frac{8 \pi W_{\mathrm{T}}}{25} \biggr) \left\{ 4
    \sinc(\pi W_{\mathrm{T}})^4 \vphantom{\sin \biggl( 4 K - \frac{9 \pi}{50}
      \biggr)} \right. \\ - \frac{1}{4} \left[ 2 \sin \biggl( 4 K - \frac{9
      \pi}{50} \biggr) \left( 1 - \sinc(\pi W_{\mathrm{T}})^2 \right)^2
  \right. \\ + \sqrt{2} \left( \sin \biggl( \frac{7 \pi}{100} \biggr) - \cos
    \biggl( \frac{7 \pi}{100} \biggr) \right) \\ \left. \left. \vphantom{\sin
        \biggl( 4 K - \frac{9 \pi}{50} \biggr)} \times \left( 1 + \sin(\pi
        W_{\mathrm{T}})^2 \right)^2 \right] \right\}^{1/2}.
\end{multline}
Upon increasing the value of $W_{\mathrm{T}}$, the expression under the square
root vanishes first when $\sin(4 K - 9 \pi/50) = -1$, which leads to
\begin{equation}
  \label{eq:W-T-c-1-K-c-1}
  W_{\mathrm{T}, \mathrm{c}, 1} \approx 0.535, \qquad K_{\mathrm{c}, 1} = \frac{21 \pi}{50},
  \frac{23 \pi}{25},
\end{equation}
i.e., the gap closes simultaneously at two points as shown in
Fig.~\ref{fig:lambda-3-4}(b). For
$W_{\mathrm{T}} > W_{\mathrm{T}, \mathrm{c}, 1}$, there are finite regions in
the Brillouin zone in which $\lambda_{3, \mathbf{K}}$ and
$\lambda_{4, \mathbf{K}}$ are purely imaginary as shown in
Figs.~\ref{fig:lambda-3-4}(c) and~(d).

For even larger values of $W_{\mathrm{T}}$, the equation
$\e^{-\imag T \lambda_{3, \mathbf{K}}} - \e^{-\imag T \lambda_{4, \mathbf{K}}} =
0$
permits further solutions that correspond to gap closings at
$\Re(T \lambda_{n, \mathbf{K}}) = \pi$, where we note that, according to
Eq.~\eqref{eq:eigenequation-F}, $\Re(T \lambda_{n, \mathbf{K}})$ is defined in
the interval $(-\pi, \pi]$. These solutions occur first when
$\sin(4 K - 9 \pi/50) = 1$. We find
\begin{equation}
  \label{eq:W-T-c-2-K-c-2}
  W_{\mathrm{T}, \mathrm{c}, 2} \approx 0.783, \qquad K_{\mathrm{c}, 2} = \frac{17 \pi}{100},
  \frac{67 \pi}{100}.
\end{equation}
The closing of the gap at $\Re(T \lambda_{n, \mathbf{K}}) = \pi$ leads to occurrence of further regions in the Brillouin
zone in which the Floquet-Bloch bands are purely imaginary as illustrated in
Figs.~\ref{fig:lambda-3-4}(e) and~(f).

\begin{figure*}
  \centering
  \includegraphics{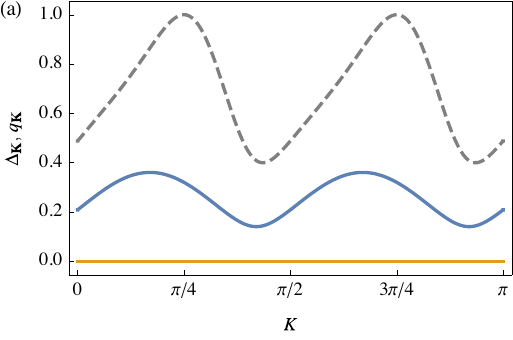}
  \includegraphics{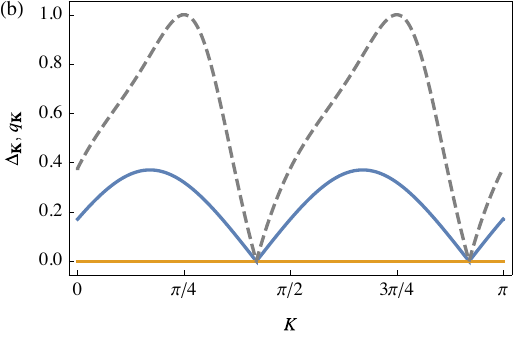} \\
  \includegraphics{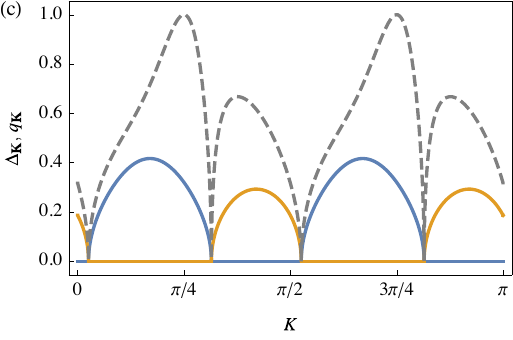}
  \includegraphics{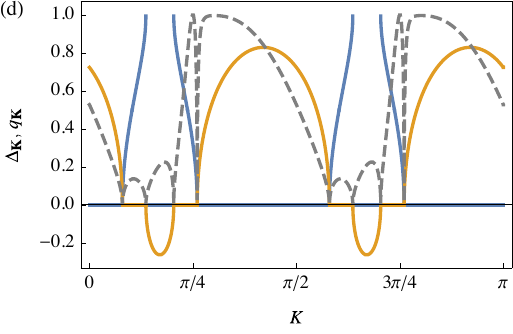}
  \caption{The real and imaginary parts of
    $\Delta_{\mathbf{K}} = T \left( \lambda_{3, \mathbf{K}} - \lambda_{4,
        \mathbf{K}} \right)/(2 \pi)$
    are shown as functions of $K = K_x = K_y$ as blue and orange lines,
    respectively. The gray dashed line shows the projection $q_{\mathbf{K}}$
    defined in Eq.~\eqref{eq:q-k}. Parameters are chosen as in
    Fig.~\ref{fig:lambda-3-4}: (a) $W_{\mathrm{T}} = 0.51 \Omega$, (b)
    $W_{\mathrm{T}} = W_{\mathrm{T, c, 1}}$, (c)
    $W_{\mathrm{T, c, 1}} < W_{\mathrm{T}} = 0.61 \Omega < W_{\mathrm{T, c,
        2}}$, (d) $W_{\mathrm{T}} = 0.8 \Omega > W_{\mathrm{T, c, 2}}$.}
  \label{fig:Delta-lambda-q}
\end{figure*}

The regions in the Brillouin zone within which the Floquet-Bloch bands
$\lambda_{n, \mathbf{K}}$ with $n = 3, 4$ are purely imaginary are bounded by
exceptional lines. At each point on these exceptional lines, both the
eigenvalues and the eigenvectors coalesce, i.e.,
$\lambda_{3, \mathbf{K}} = \lambda_{4, \mathbf{K}}$ and
$\kket{\rho_{3, \mathbf{K}}} = \kket{\rho_{4, \mathbf{K}}}$. To illustrate these
properties, we introduce the projector onto the subspace that is orthogonal to
$\kket{\rho_{3, \mathbf{K}}}$,
\begin{equation}
  \mathcal{Q}_{3, \mathbf{K}} = \id - \kket{\rho_{3, \mathbf{K}}}
  \bbra{\rho_{3, \mathbf{K}}},
\end{equation}
and we consider the projection of $\kket{\rho_{4, \mathbf{K}}}$,
\begin{equation}
  \label{eq:q-k}
  q_{\mathbf{K}} = \norm{\mathcal{Q}_{3, \mathbf{K}} \kket{\rho_{4, \mathbf{K}}}}.
\end{equation}
When $q_{\mathbf{K}}$ vanishes, the eigenvectors $\kket{\rho_{3, \mathbf{K}}}$
and $\kket{\rho_{4, \mathbf{K}}}$ coalesce. The projection $q_{\mathbf{K}}$,
along with the normalized difference of eigenvalues
$\Delta_{\mathbf{K}} = T \left( \lambda_{3, \mathbf{K}} - \lambda_{4,
    \mathbf{K}} \right)/(2 \pi)$,
is shown in Fig.~\ref{fig:Delta-lambda-q}. In particular, the figure shows cuts
through the Brillouin zone for $K = K_x = K_y$. Exceptional points with
$\Delta_{\mathbf{K}} = 0$ appear first for
$W_{\mathrm{T}} > W_{\mathrm{T}, \mathrm{c}, 1}$ at the intersection of the
Brillouin-zone cut and the exceptional line. Further exceptional points for
which $\Delta_{\mathbf{K}} = 1$ appear when $W_{\mathrm{T}} > W_{\mathrm{T, c, 2}}$.

\section{Topological invariant}
\label{sec:topol-invar}

To understand why the closing of a bulk gap may be related to a topological
phase transition induced by temporal noise, we attempt to generalize the topological
winding number of Ref.~\cite{Rudner2013} to this nonunitary case. As a
reminder, the unitary winding number is given by
\begin{equation}
  \label{eq:W-U}
  W_{\varepsilon} = \int \frac{\mathrm{d} t \, \mathrm{d}^2 \mathbf{k}}{8 \pi^2}
  \tr(U_{\varepsilon}^{\dagger} \partial_t U_{\varepsilon} [U_{\varepsilon}^{\dagger} \partial_{k_x} U_{\varepsilon},
  U_{\varepsilon}^{\dagger} \partial_{k_y} U_{\varepsilon}]).
\end{equation}
where the integration extends over
$\left( t, \mathbf{k} \right) \in [0, T] \times \mathrm{BZ}$. For the AFAI, the
Brillouin Zone $\mathrm{BZ}$ is the square with vertices
$\left( -\pi, 0 \right)$, $\left( 0, \pi \right)$, $\left( \pi, 0 \right)$,
$\left( 0, - \pi \right)$. This can be seen by noting that the AFAI is defined on a 2D lattice with
primitive lattice vectors
\begin{equation}
  \mathbf{a}_1 = \left( 2, 0 \right), \qquad \mathbf{a}_2 = \left( 1, 1
  \right).
\end{equation}
The corresponding reciprocal lattice is spanned by the vectors
\begin{equation}
  \begin{split}
    \mathbf{b}_1 & = 2 \pi \frac{R \mathbf{a}_2}{\mathbf{a}_1 \cdot R
      \mathbf{a}_2} = \pi \left( 1, - 1 \right), \\ \mathbf{b}_2 & =
    2 \pi \frac{R \mathbf{a}_1}{\mathbf{a}_2 \cdot R \mathbf{a}_1} = 2 \pi \left( 0, 1 \right),
  \end{split}
\end{equation}
where the matrix $R$ describes a rotation by $\pi/2$,
\begin{equation}
  R =
  \begin{pmatrix}
    0 & 1 \\ -1 & 0
  \end{pmatrix}.
\end{equation}
The Brillouin Zone $\mathrm{BZ}$ is given by the region that is bounded by
straight lines that intersect the reciprocal basis vectors halfway. This
construction leads to the square with vertices specified above. However, due to the symmetry of the integrand in
Eq.~\eqref{eq:W-U} with respect to translations by the reciprocal lattice
vectors $\pm \pi \left( 1, 1 \right) = \pm \left( \mathbf{b}_1 + \mathbf{b}_2 \right)$, the ranges of integration for $k_x$ and
$k_y$ can be taken to be $k_x \in [0, \pi)$ and $k_y \in [-\pi, \pi)$.

The unitary operator $U_{\varepsilon} = U_{\varepsilon, \mathbf{k}}(t)$ is
defined as
\begin{equation}
  \label{eq:U-epsilon}
  U_{\varepsilon, \mathbf{k}}(t) =
  \begin{cases}
    U_{\mathbf{k}}(2 t) & \text{for } 0 \leq t < T/2, \\
    V_{\varepsilon, \mathbf{k}}(2 \left( T - t \right)) & \text{for } T/2 \leq t
    < T,
  \end{cases}
\end{equation}
where $U_{\mathbf{k}}(t)$ is the time evolution operator for momentum mode
$\mathbf{k}$ such that $U_{\mathbf{k}}(T) = U_{\mathbf{k}}$ is the Floquet
operator, and the trivial ``return map'' $V_{\varepsilon, \mathbf{k}}(t)$ is given by
\begin{equation}
  \label{eq:V-epsilon}
  V_{\varepsilon, \mathbf{k}}(t) = \e^{-\imag t H_{\mathrm{eff}, \varepsilon,
      \mathbf{k}}}, \qquad H_{\mathrm{eff}, \varepsilon, \mathbf{k}} =
  \frac{\imag}{T} \ln(U_{\mathbf{k}}).
\end{equation}
Here, the branch cut of the logarithm is chosen to lie along the direction
$\e^{-\imag \varepsilon}$. For the choices $\varepsilon = 0$ and
$\varepsilon = \pi$, the winding numbers $W_0$ and $W_{\pi}$ detect the presence
of edge modes at quasienergies $T \varepsilon_{n, \mathbf{k}} = 0$ and
$T \varepsilon_{n, \mathbf{k}} = \pi$, respectively.

We generalize the definition of $W_{\varepsilon}$ in Eq.~\eqref{eq:W-U} in two
steps: First, we consider the case of a unitary Floquet superoperator, which we
obtain in the limit $W_{\mathrm{T}} = 0$, and which is given in
Eq.~\eqref{eq:F-unitary}. According to the latter equation, $U_{\varepsilon}$ in
Eq.~\eqref{eq:W-U} should be replaced by
\begin{equation}
  \label{eq:cal-U}
  \mathcal{U}_{\varepsilon, \mathbf{k}_1, \mathbf{k}_2}(t) = U_{\varepsilon,
    \mathbf{k}_2}^{*}(t) \otimes U_{\varepsilon, \mathbf{k}_1}(t).
\end{equation}
We require that the topological properties of $U_{\varepsilon}$ and
$\mathcal{U}_{\varepsilon}$ are identical. This condition is satisfied
by the following definition of a topological invariant for
$\mathcal{U}_{\varepsilon}$:
\begin{multline}
  \label{eq:W-cal-U}
  W_{\varepsilon} = \int \frac{\mathrm{d} t \, \mathrm{d}^2 \mathbf{k}_1 \,
    \mathrm{d}^2 \mathbf{k}_2}{32 \pi^2 v_{\mathrm{BZ}}} \tr \! \left(
    \mathcal{U}_{\varepsilon}^{\dagger} \partial_t \mathcal{U}_{\varepsilon} \right. \\
  \left. \times ([\mathcal{U}_{\varepsilon}^{\dagger} \partial_{k_{1, x}} \mathcal{U}_{\varepsilon},
    \mathcal{U}_{\varepsilon}^{\dagger} \partial_{k_{1, y}} \mathcal{U}_{\varepsilon}] +
    [\mathcal{U}_{\varepsilon}^{\dagger} \partial_{k_{2, x}} \mathcal{U}_{\varepsilon},
    \mathcal{U}_{\varepsilon}^{\dagger} \partial_{k_{2, y}} \mathcal{U}_{\varepsilon}]) \right).
\end{multline}
It is straightforward to check that due to the factorization of
$\mathcal{U}_{\varepsilon}$ in Eq.~\eqref{eq:cal-U}, the first term in the
second line of Eq.~\eqref{eq:W-cal-U} does not depend on
$\mathbf{k}_2$. Therefore, for this term, the integration over $\mathbf{k}_2$
simply yields a factor $v_{\mathrm{BZ}} = 2 \pi^2$, which is the volume of the
Brillouin zone. Further, the multiplication by $1/4$ in comparison to
Eq.~\eqref{eq:W-U} accounts for a factor of $\tr(\id) = 2$, and for the double
counting due to the last term in Eq.~\eqref{eq:W-cal-U}, which symmetrizes the
definition of $W_{\varepsilon}$ with respect to $\mathbf{k}_1$ and
$\mathbf{k}_2$. Consequently, the definitions in Eqs.~\eqref{eq:W-U}
and~\eqref{eq:W-cal-U} yield the same values of the winding number.

As a second step, we replace $\mathcal{U}_{\varepsilon}$ by a suitably chosen
$\mathcal{F}_{\varepsilon}$. To this end, we have to give meaning to
$\mathcal{F}_{\mathbf{k}_1, \mathbf{k}_2}(t)$. An obvious choice is given by
\begin{equation}
  \mathcal{F}_{\mathbf{k}_1, \mathbf{k}_2}(t) = \overline{\mathcal{U}_{\mathbf{k}_1, \mathbf{k}_2}(t)},
\end{equation}
where the overline denotes an average over $\delta_m$ as in the definition of
the Floquet superoperator in Eq.~\eqref{eq:F}. However, this is slightly
difficult to work with if we think of $W_{\mathrm{T}}$ as timing noise. Instead,
we make the equivalent choice to think of
$\delta_m \in [-W_{\mathrm{T}}, W_{\mathrm{T}}]$ not as timing noise but as a
random coefficient of the Hamiltonian. Then, at time $t$ with
$\left( m - 1 \right) T/M \leq t < m T/M$ where $m = 1, \dotsc, M$, we obtain
\begin{multline}
  \mathcal{F}_{\mathbf{k}_1, \mathbf{k}_2}(t) = \mathcal{U}_{\mathbf{k}_1,
    \mathbf{k}_2, m}(t) \mathcal{E}_{\mathbf{k}_1, \mathbf{k}_2, m}(t) \\ \times
  \mathcal{F}_{\mathbf{k}_1, \mathbf{k}_2, m - 1} \mathcal{F}_{\mathbf{k}_1,
    \mathbf{k}_2, m - 2} \dotsb \mathcal{F}_{\mathbf{k}_1, \mathbf{k}_2, 1},
\end{multline}
where
\begin{equation}
  \begin{split}
    \mathcal{U}_{\mathbf{k}_1, \mathbf{k}_2, m}(t) & = \e^{- \imag \left[ t -
        \left( m - 1
        \right) T/M \right] \mathcal{H}_{\mathbf{k}_1, \mathbf{k}_2, m}}, \\
    \mathcal{E}_{\mathbf{k}_1, \mathbf{k}_2, m}(t) & = \sinc( \left[ t - \left(
        m - 1 \right) T/M \right] W_{\mathrm{T}} \mathcal{H}_{\mathbf{k}_1,
      \mathbf{k}_2, m} ), \\ \mathcal{F}_{\mathbf{k}_1, \mathbf{k}_2, m} & =
    \mathcal{U}_{\mathbf{k}_1, \mathbf{k}_2, m} \mathcal{E}_{\mathbf{k}_1,
      \mathbf{k}_2, m}
  \end{split}
\end{equation}
and
\begin{equation}
  \begin{split}
    \mathcal{U}_{\mathbf{k}_1, \mathbf{k}_2, m} & = \mathcal{U}_{\mathbf{k}_1,
      \mathbf{k}_2, m}(m T/M), \\ \mathcal{E}_{\mathbf{k}_1, \mathbf{k}_2, m} &
    = \mathcal{E}_{\mathbf{k}_1, \mathbf{k}_2, m}(m T/M).
  \end{split}
\end{equation}
With this definition of $\mathcal{F}_{\mathbf{k}_1, \mathbf{k}_2}(t)$, we can
generalize Eq.~\eqref{eq:U-epsilon} as
\begin{equation}
  \label{eq:F-epsilon}
  \mathcal{F}_{\varepsilon, \mathbf{k}_1, \mathbf{k}_2}(t) =
  \begin{cases}
    \mathcal{F}_{\mathbf{k}_1, \mathbf{k}_2}(2 t) & \text{for } 0 \leq t < T/2, \\
    \mathcal{G}_{\varepsilon, \mathbf{k}_1, \mathbf{k}_2}(2 \left( T - t \right)) & \text{for }
    T/2 \leq t < T,
  \end{cases}
\end{equation}
where
\begin{equation}
  \label{eq:cal-G}
  \begin{split}
    \mathcal{G}_{\varepsilon, \mathbf{k}_1, \mathbf{k}_2}(t) & = \e^{- \imag t
      \mathcal{H}_{\mathrm{eff}, \varepsilon, \mathbf{k}_1, \mathbf{k}_2}}, \\
    \mathcal{H}_{\mathrm{eff}, \varepsilon, \mathbf{k}_1, \mathbf{k}_2} & =
    \frac{\imag}{T} \ln(\mathcal{F}_{\mathbf{k}_1, \mathbf{k}_2}).
  \end{split}
\end{equation}
We set
$\mathcal{F}_{\mathbf{k}_1, \mathbf{k}_2} = \mathcal{F}_{\mathbf{k}_1,
  \mathbf{k}_2}(T)$,
and as above we choose the branch cut of the logarithm to lie along the
direction $\e^{-\imag \varepsilon}$. A winding number of the Floquet
superoperator is thus given by
\begin{multline}
  \label{eq:W-cal-F}
  W_{\varepsilon} = \int \frac{\mathrm{d} t \, \mathrm{d}^2 \mathbf{k}_1 \,
    \mathrm{d}^2 \mathbf{k}_2}{32 \pi^2 v_{\mathrm{BZ}}} \tr \! \left(
    \mathcal{F}_{\varepsilon}^{-1} \partial_t \mathcal{F}_{\varepsilon} \right. \\
  \left. \times ([\mathcal{F}_{\varepsilon}^{-1} \partial_{k_{1, x}}
    \mathcal{F}_{\varepsilon},
    \mathcal{F}_{\varepsilon}^{-1} \partial_{k_{1, y}}
    \mathcal{F}_{\varepsilon}] \right. \\ \left. +
    [\mathcal{F}_{\varepsilon}^{-1} \partial_{k_{2, x}}
    \mathcal{F}_{\varepsilon},
    \mathcal{F}_{\varepsilon}^{-1} \partial_{k_{2, y}}
    \mathcal{F}_{\varepsilon}]) \right),
\end{multline}
where, in comparison to Eq.~\eqref{eq:W-cal-U}, $\mathcal{U}_{\varepsilon}$ is
replaced by $\mathcal{F}_{\varepsilon}$. Further, to account for the fact that
the Floquet superoperator is, in general, not unitary, we replace the Hermitian
conjugate $\mathcal{U}_{\varepsilon}^{\dagger}$ by the inverse
$\mathcal{F}_{\varepsilon}^{-1}$.

\begin{figure}
  \centering
  \includegraphics{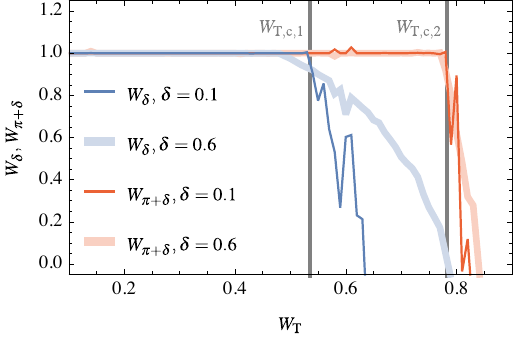}
  \caption{Winding numbers $W_{\delta}$ and $W_{\pi + \delta}$. Topological
    transitions occur at $W_{\mathrm{T, c, 1}}$ and $W_{\mathrm{T, c, 2}}$. We
    introduce the regularization $\delta$ to avoid ill-defined values of the
    logarithm in Eq.~\eqref{eq:cal-G}. Larger values of $\delta$ improve the
    results of the numerical integration, but also lead to a rounding of the
    topological transitions.}
  \label{fig:topological-invariant}
\end{figure}

Results for the winding number $W_{\varepsilon}$ of $\mathcal{F}_{\varepsilon}$
are shown in Fig.~\ref{fig:topological-invariant} for
$\varepsilon = \delta, \pi + \delta$, where we introduce $0 < \delta < 1$ to
regularize the numerical integration. This regularization is required for
$W_{\delta}$ when $W_{\mathrm{T}} > W_{\mathrm{T, c, 1}}$. Then, finite portions
of the bands $\lambda_{3, \mathbf{K}}$ and $\lambda_{4, \mathbf{K}}$, which are
shown as blue and red lines in Fig.~\ref{fig:F-spectrum} and which appear to
encode the topology of $\mathcal{F}_{\varepsilon}$, are located on the positive
real axis in the complex $\e^{-\imag T \lambda}$ plane, rendering the logarithm
with the branch cut along $\e^{-\imag \varepsilon}$ with $\varepsilon = 0$
ill-defined. Similarly, $W_{\pi + \delta}$ requires regularization when
$W_{\mathrm{T}} > W_{\mathrm{T, c, 2}}$. Larger values of $\delta$ lead to
improved convergence of the numerical integration. However, at the same time,
the transitions at $W_{\mathrm{T, c, 1}}$ and $W_{\mathrm{T, c, 2}}$ become
washed out.

For $W_{\mathrm{T}} < W_{\mathrm{T, c, 1}}$, we find
\begin{equation}
  W_{\delta} = W_{\pi + \delta} = 1.
\end{equation}
In contrast, for $W_{\mathrm{T}} > W_{\mathrm{T}, \mathrm{c}, 1}$, the winding
number $W_{\delta}$ is not quantized because of the dispersing modes sitting on
the real axis in the complex $\e^{-\imag T \lambda}$ plane. Therefore, we see a
topological transition at $W_{\mathrm{T, c, 1}} \approx 0.535$, which is
consistent with the value obtained in Fig.~4 of the main text through entirely
different means.

\bibliography{dissipative_AFAI}
